\documentclass[a4paper,11pt]{article}
\usepackage[utf8]{inputenc}
\usepackage[T1]{fontenc}

\usepackage{booktabs}
\usepackage{subcaption}
\usepackage[table]{xcolor}
\usepackage{xcolor}

\usepackage{graphicx}
\usepackage{subcaption}

\usepackage{amsmath}
\usepackage{amssymb}
\usepackage{empheq}

\definecolor{lightgray}{gray}{0.95}
\definecolor{lightgrayB}{gray}{0.85}

\newcommand{\NbmodeP}[0]{N_p}
\newcommand{\NbmodeU}[0]{N_u}

\newcommand{\fract}[2]{\displaystyle{\frac{#1}{#2}}}

\newcommand{\vit}{\mathbf{u}}
\newcommand{\param}{Re}
\newcommand{\press}{p}

\newcommand{\vbasis}{\boldsymbol{\varphi}}
\newcommand{\pbasis}{\psi}

\newcommand{\vcoef}{a}
\newcommand{\pcoef}{b}

\title{Non intrusive parametric Reduced Order Models based on physics for incompressible flow}
\author{Cyrille Allery, Claudine Béghein\footnote{Corresponding author}}

\begin{document}
	\maketitle
\begin{center}
	LaSIE, UMR CNRS 7356, La Rochelle Université\\
	Av. Michel Crépeau, 17042 La Rochelle Cedex 1, France\\
	\{callery, cbeghein\}@univ-lr.fr\\
\end{center}

\section*{Abstract}
Reduced order modeling provides an efficient alternative to high fidelity simulations for the rapid prediction of fluid flows in many query settings. Among non intrusive approaches, Operator Inference (OpInf) enables the identification of reduced dynamical operators directly from data while preserving a prescribed mathematical structure. The present work extends physics structured OpInf to the construction of parametric Reduced Order Models for transient incompressible flows governed by the Navier-Stokes equations. The velocity and pressure fields are represented using separate global reduced bases, and their reduced operators are identified directly from temporal and parametric POD coefficients. The identification procedure is formulated as a sequence of uncoupled least squares problems, leading to an efficient offline stage while avoiding the explicit projection of the governing equations. The approach is first evaluated on a lid driven cavity flow in a periodic regime to assess temporal extrapolation and prediction at previously unseen parameter values. Its applicability is then demonstrated for transient flow through a porous medium with a square obstacle, for which the nonlinear terms are not directly amenable to a standard Galerkin projection. The results demonstrate that the approach can accurately reproduce both velocity and pressure fields while providing computationally efficient predictions for unseen parameter values.\\

\noindent \textbf{Keywords : }reduced order models, Proper Orthogonal Deaomposition, incompressible flows , Navier-Stokes equations.

\section{Introduction}
Fluid flows are encountered in many engineering and environmental applications, such as aerodynamic design, heat transfer, groundwater transport, filtration, energy systems, and flow through porous media. Their numerical simulation is typically performed using high fidelity methods, including finite element, finite volume, or finite difference discretizations. Although these approaches provide accurate solutions, they often involve a large number of degrees of freedom, resulting in substantial computational costs and memory requirements. Such computational demands become prohibitive in many query contexts, where the governing equations must be solved repeatedly for different operating conditions, geometrical configurations, or physical parameters. This is the case, for example, in optimization, uncertainty quantification, parameter estimation, digital twins, and real time monitoring or control.

Reduced order modeling offers an efficient way to address this issue by significantly reducing the computational cost while retaining the main features of the flow dynamics. The general idea is to approximate the high dimensional solution in a low dimensional subspace spanned by a small number of spatial basis functions. Once this reduced basis is available, the evolution of the system in time and/or with respect to parameters can be computed using a reduced order model (ROM). A wide range of reduced order modeling techniques has been developed to construct low dimensional approximation spaces. These methods can be broadly classified into {\it a posteriori} linear approaches, including Proper Orthogonal Decomposition (POD)\cite{Sirovich87}, Reduced Basis method \cite{Buffa2012} or Dynamic Mode Decomposition\cite{Schmid2010}; {\it a priori} linear approaches, such as A Priori Reduction method \cite{Ryckelynck2005}\cite{Verdon2011} or Proper Generalized Decomposition method \cite{Ammar2007}\cite{Dumon2011}; and nonlinear manifold learning techniques, including kernel PCA method \cite{Scholkopf98}, Locally Linear Embedding method \cite{Roweis2000} \ldots. Among these techniques, the POD remains the most widely adopted in fluid dynamics. It constructs a reduced basis from a collection of representative high fidelity solutions by extracting the dominant coherent structures of the flow. The resulting spatial basis is optimal in the sense that it minimizes the projection error for a prescribed number of modes. Once this reduced spatial basis has been constructed, the dynamics of the system are entirely characterized by the evolution of the associated reduced coefficients. Methods for predicting these coefficients under new operating conditions are generally divided into two families: intrusive and non intrusive approaches.

In intrusive ROMs, the governing equations are projected onto the reduced basis, most
commonly using a Galerkin projection, thereby preserving the mathematical structure of
the original system. Galerkin POD-ROMs have been successfully applied in many fields,
including fluid mechanics \cite{rowley_2017,beghein_2014}, flow control
\cite{tallet_2016,borggaard_2015,oulghelou_2022}, structural dynamics
\cite{touze_2016}, thermal engineering \cite{Xu2020,ZHAO2026114781}, and
fluid structure interaction \cite{Liberge_2010,Falaize_2019}. These studies have
shown that Galerkin POD/ROMs can accurately reproduce transient behaviours while
reducing the computational cost by several orders of magnitude. Nevertheless, their
construction generally requires access to the discretized governing equations.
Furthermore, stabilization procedures may be required to ensure robustness for
convection dominated or strongly nonlinear problems.

In contrast, non intrusive ROMs learn the reduced dynamics directly from data
without requiring access to the governing equations. A common approach consists of
predicting the reduced solution for new parameter values using interpolation or
regression techniques. In the simplest case, a single global reduced basis is
constructed from snapshots collected over the training set, and only the associated
reduced coefficients are interpolated for previously unseen operating conditions.
More advanced methods also interpolate the reduced basis itself. For example, the
Bi-CITSGM (Bi-Calibrated Interpolation on Tangent Subspaces of the Grassmann
Manifold) method interpolates both the reduced coefficients and the spatial basis,
allowing the reduced approximation space to vary with the parameters
\cite{oulghelou_2020,Dubot2021}. Machine learning techniques have also been widely
used to predict reduced coefficients. Multilayer perceptrons are among the most
common choices \cite{Hesthaven2018,Pawar2019,DiasRibeiro2023}, while other regression
methods, such as Gaussian Process Regression, have also been investigated
\cite{Demo2021,Kadeethum2021}. Since these approaches rely solely on data, they do
not automatically satisfy the underlying physical equations. Physics Informed Neural
Networks (PINNs) address this limitation by incorporating the governing equations
into the training process \cite{Raissi2019,Chen2021,Hijazi2023}.

Rather than directly predicting the reduced coefficients, another line of research focuses on identifying the reduced dynamical system governing their temporal evolution. Several data driven approaches have been developed for this purpose. Sparse identification methods, such as SINDy \cite{brunton_2016,rudy_2017,kaheman_2020}, assume that the underlying dynamics can be represented as a sparse combination of candidate nonlinear functions. A dictionary of candidate terms, typically composed of polynomial and trigonometric functions, is first constructed, and sparse regression is then employed to identify a parsimonious subset of active terms that best describes the dynamics. Modal identification methods \cite{girault_2004,balima_2006,rouizy_2010,rouizy_2013} instead determine the reduced order coefficients by minimizing the discrepancy between the ROM predictions and the corresponding high fidelity solutions. Although accurate, the associated optimization procedure may become computationally expensive. Polynomial surrogate models \cite{brillaut_2025} approximate the reduced dynamics through explicit polynomial mappings of the reduced variables. Finally, other approaches aim at preserving the algebraic structure of the governing equations. For instance, the Lift $\&$ Learn framework \cite{QIAN2020} introduces appropriate lifted variables to identify reduced operators while retaining the structure of the underlying dynamics, whereas DeepHyROMnet \cite{Cicci2022} combines deep neural networks with hyper reduction techniques for nonlinear parametric partial differential equations.

Among these approaches, Operator Inference (OpInf) provides a particularly attractive framework for learning reduced dynamical systems while retaining a prescribed mathematical structure. Originally introduced in \cite{Peherstorfer2016}, OpInf consists in directly inferring the reduced operators governing the dynamics from data. Given the reduced coefficients, these operators are identified through a regression problem involving the reduced state snapshots and their time derivatives. The structure of the reduced model is prescribed according to the underlying physics and may include linear, polynomial, or input dependent operators. OpInf has been investigated for a wide range of scientific and engineering applications, including reactive flows and combustion, additive manufacturing, structural and mechanical systems, and multiphysics energy systems. A comprehensive overview of these developments is provided in \cite{Kramer2024}.

Several studies have focused specifically on incompressible flows, for which the pressure field raises additional difficulties. Among the recent extensions of OpInf, distributed Operator Inference \cite{Farcas2025} provides a parallel algorithm for efficiently constructing physics based reduced order models for problems with large state dimensions. It was applied to twodimensional Navier-Stokes flows, but the considered formulation focused on the identification of the velocity dynamics and did not address parametric dependence. A different strategy \cite{Koike2026} combines the velocity and pressure coefficients into a single POD reduced state and identifies them through a quadratic reduced model whose form is not directly derived from the Navier-Stokes equations. The method was validated on benchmark problems and on a large scale turbulent channel flow, but in a purely temporal setting. Similarly, \cite{Rocha2023} applied POD to a vector collecting the velocity components and the pressure. The predictive capabilities of the model were assessed on two test cases at higher Rayleigh numbers, without considering parametric dependence. In both works, gathering all variables into a single vector leads to reduced bases that lack a direct physical interpretation and may be suboptimal for the representation of individual variables.

A different strategy \cite{Benner2022} exploits the structure of the incompressible Navier-Stokes equations to infer reduced models that recover both the velocity and pressure fields, the latter being required in many applications of practical interest. In this approach, the velocity and pressure dynamics are decoupled during the inference procedure, yielding an efficient non intrusive reduced order model. The method was assessed on two canonical two dimensional problems, namely the lid driven cavity flow and the flow past a cylinder in a stationary regime. It was shown to reproduce the dominant flow behavior while providing an efficient alternative to intrusive projection based reduction. It should be noted, however, that the inferred dynamics remain valid only for times slightly beyond the sampling window and that no parametric dependence is considered.

Parametric OpInf was considered in \cite{Kang2026}, where a low dimensional representation of a high fidelity model of the purging process in semiconductor manufacturing was constructed for varying argon mass flow rates and outlet pressures. The approach learns reduced models at several parameter values and subsequently interpolates them to predict the system behavior at previously unseen parameter combinations. In this case, however, the reduced operators are not global, but are constructed separately for different parameter values and then interpolated, which limits the flexibility of the parametric model. Another extension of OpInf is the probabilistic formulation proposed in \cite{Yong2025}, which quantifies model form uncertainties by randomizing the projection matrix and expanding the approximation space. It was assessed on the Burgers and Navier-Stokes equations.

Still in the context of fluid flows, another approach called GILD \cite{oulghelou_2024} was proposed for parametric flows. It assumes a prescribed quadratic reduced model consistent with the Galerkin projection of the Navier-Stokes equations and identifies its coefficients through a greedy optimization strategy based on residual minimization. The I-GILD extension \cite{oulghelou_2025} further improves computational efficiency by reformulating the identification problem in matrix form and minimizing a Frobenius norm objective.

The present work builds on these developments while addressing their main limitations. It is developed for the construction of parametric ROMs of transient incompressible flows governed by the Navier-Stokes equations, with the objective of providing an efficient and robust framework for the rapid prediction of velocity and pressure fields at previously unseen parameter values. In contrast with \cite{Benner2022}, where no parametric dependence is considered, the methodology considered here extends physics structured OpInf to parametric problems while recovering both the velocity and pressure fields. Contrary to the GILD and I-GILD methodologies \cite{oulghelou_2024,oulghelou_2025}, the pressure is explicitly taken into account and a single global reduced basis is used instead of local bases that are subsequently interpolated, which considerably simplifies the overall implementation. Moreover, the identification procedure is formulated as a sequence of uncoupled least squares problems, making the offline stage particularly efficient. Unlike Galerkin approaches, the reduced operators are identified directly from the temporal and parametric POD coefficients, thereby avoiding the explicit projection of the governing equations. As a result, the considered methodology can be applied to systems for which Galerkin projection is difficult or impractical, such as flows in porous media, where nonlinear terms such as the Forchheimer contribution cannot be projected onto the reduced basis \cite{allery_beghein_2025}. By identifying the reduced operators directly from data, the considered methodology naturally overcomes this limitation. The approach is illustrated on two representative parametric benchmark problems, namely the lid driven cavity flow in a periodic regime and transient flows in a porous medium with a square obstacle, highlighting both the generality of the methodology and its ability to construct accurate and computationally efficient parametric ROMs.
~\\

The remainder of this paper is organized as follows. Section 2 briefly recalls the POD-Galerkin ROM formulation for the incompressible Navier-Stokes equations,
including the velocity and pressure models, which provides the reference framework for the considered approach. Section 3 presents the non intrusive reduced order modeling strategy based on the equations and describes the identification procedures used to construct the velocity and pressure reduced models. Section 4 extends the methodology to transient flows in porous media.  The performance of the proposed approach is first aplied to the lid driven cavity flow. Its ability to extrapolate the solution in time beyond the interval used for training is initially investigated for a fixed parameter value. The methodology is then evaluated in a parametric setting to predict flow fields at previously unseen parameter values. Finally, the approach is applied to porous media flows to assess its applicability to this class of problems.

\section{Navier–Stokes POD Galerkin ROM}
\subsection{Non dimensional incompressible Navier–Stokes equations}
Let $\Omega \subset \mathbb{R}^d \ (d=2,3$) be a bounded domain with boundary
$\partial\Omega = \Gamma_D \cup \Gamma_N$ such as $\Gamma_D \cap \Gamma_N = \varnothing$,
where $\Gamma_D$ and $\Gamma_N$ denote the Dirichlet and Neumann parts of the boundary, respectively.
The parametrized non dimensionnal incompressible Navier-Stokes problem consists in finding, for a given Reynolds number, the non dimensionnal velocity field $\vit : \Omega \times (0,T) \to \mathbb{R}^d $ and the non dimensionnal pressure $\press : \Omega \times (0,T) \to \mathbb{R}$,
such that
\begin{empheq}[left=\empheqlbrace]{align}
\nabla\cdot\vit
&= 0
\qquad \text{in } \Omega\times(0,T).\\
\partial_t \vit + (\vit\cdot\nabla)\vit - \fract{1}{Re}\Delta\vit +\nabla\press
&= \mathbf{f}
\qquad \text{in } \Omega\times(0,T), \label{equ:momentum}\\
\vit &= \vit_D
\qquad \text{on } \Gamma_D\times(0,T),
\\
\bigl(-\press \mathbf{I}
+ \mu(\nabla\vit + \nabla\vit^{T})\bigr)\mathbf{n}
&= \mathbf{g}_N \qquad \text{on } \Gamma_N\times(0,T),\\
\vit(\mathbf{x},0)&=\vit_0(\mathbf{x})
\qquad \text{in } \Omega
\end{empheq}
where $Re=\fract{U_{ref}L_{ref}}{\nu}$ is the Reynolds number. $\nu$ denotes the kinematic viscosity, $\mu$ the dynamic viscosity, $U_{ref}$ a reference velocity, $L_{ref}$ a reference length, $\mathbf{n}$ the outward unit normal vector to $\partial\Omega$ and $\mathbf{f}$ is a given body force.
The initial condition $\vit(\mathbf{x},0)$ satisfies the boundary conditions and is divergence free.

\subsection{POD Galerkin ROM for the velocity}
Starting from the incompressible Navier-Stokes equations introduced in the previous section, a ROM is obtained by projecting the governing equations onto a low dimensional space generated by Proper Orthogonal Decomposition (POD) modes. The reduced velocity and pressure fields are approximated by the truncated POD expansions
\begin{eqnarray}
\vit(\mathbf{x},t,\param) \simeq \vit_r(\mathbf{x},t,\param)=
\sum_{i=1}^{\NbmodeU}
\vcoef_i(t,\param)\,
\vbasis_i(\mathbf{x}),
\label{eq:velicity_expansion}
\\
\press(\mathbf{x},t,\param)
\simeq
\press_r(\mathbf{x},t,\param)
=\sum_{k=1}^{\NbmodeP}
\pcoef_k(t,\param)\,
\pbasis_k(\mathbf{x}),
\label{eq:pressure_expansion}
\end{eqnarray}
where $\{\vbasis_i\}_{i=1}^{\NbmodeU}$ and $\{\pbasis_k\}_{k=1}^{\NbmodeP}$ denote the POD bases associated with the velocity and pressure fields, respectively. The quantities $\vcoef_i(t,\param)$ and $\pcoef_k(t,\param)$ represent the reduced coordinates describing the temporal and parametric evolution of the solution.
Substituting the approximations \eqref{eq:velicity_expansion} and \eqref{eq:pressure_expansion} into the momentum equation \eqref{equ:momentum} and performing a Galerkin projection
onto the reduced velocity space
$V_r=\mathrm{span}
\left\{
\vbasis_1,\ldots,\vbasis_{\NbmodeU}
\right\}
$ we obtain
\begin{equation}
\left(
\partial_t \vit_r,
\vbasis_i
\right)
+
\fract{1}{Re}
\left(
\nabla \vit_r,
\nabla \vbasis_i
\right)
+
\left(
(\vit_r \cdot \nabla)\vit_r,
\vbasis_i
\right)
=
\left(
\mathbf{f},
\vbasis_i
\right)-(
\nabla \press_r,
\vbasis_i
).
\label{eq:galerkin_projection}
\end{equation}
where $(\bullet,\bullet)$ designs the inner product in $L^2(\Omega)$.
Because the velocity basis functions satisfy the incompressibility constraint, the projected pressure is given by
\begin{equation}
(
\nabla \press_r,
\vbasis_i
)
=
-
\underbrace{(
	\press_r,
	\nabla \cdot \vbasis_i
	)}_{=0}
+
\int_{\partial \Omega}
\press_r
\,
\vbasis_i \cdot \mathbf{n}
\, ds,
\end{equation}
This term vanishes exactly when the velocity field, and consequently the basis functions $\vbasis_i$, satisfy homogeneous Dirichlet boundary conditions. For many flows, the pressure term is approximated on each boundary by its time-averaged value $\overline{p}(\mathbf{x})$, so that
\begin{equation}
(
\nabla \press_r,
\vbasis_i
)
\simeq
\int_{\partial \Omega}
\overline{p}
\,
\vbasis_i \cdot \mathbf{n}
\, ds.
\end{equation}
After substituting the reduced velocity expansion \eqref{eq:velicity_expansion} into \eqref{eq:galerkin_projection}, and taking into account the orthonormality of the basis functions  $\vbasis_i$   the following system of ordinary differential equations, that constitutes the ROM is obtained:
\begin{equation}
\fract{d \vcoef_i}{dt}
+
\sum_{j=1}^{\NbmodeU}
\sum_{k=1}^{\NbmodeU}
Q_{ijk}
\vcoef_j
\vcoef_k+\fract{1}{Re}
\sum_{j=1}^{\NbmodeU}
L_{ij}
\vcoef_j
+
F_i=0,
\qquad
i=1,\ldots,\NbmodeU.
\label{eq:rom_system}
\end{equation}
where the reduced operators are defined by
\begin{equation*}
\begin{array}{ccc}
Q_{ijk}
=
\Big(
(\vbasis_j \cdot \nabla)\vbasis_k,
\vbasis_i
\Big),&
L_{ij}
=
(
\nabla \vbasis_j,
\nabla \vbasis_i
),&  F_i
=
-(
\mathbf{f},
\vbasis_i
)+\displaystyle{\int_{\partial \Omega}}
\overline{p}
\,
\vbasis_i \cdot \mathbf{n}
\, ds,.
\end{array}
\end{equation*}
Equation \eqref{eq:rom_system} constitutes the POD Galerkin ROM for the incompressible Navier-Stokes equations. The original partial differential equations are thus transformed into a low dimensional nonlinear dynamical system governing the evolution of the reduced coefficients $\vcoef_i(t,\param)$.

\subsection{POD Galerkin ROM for the pressure}
Although the pressure field does not explicitly appear in the reduced momentum equations, it can be reconstructed a posteriori by solving a Poisson equation.\footnote{It is also possible to construct ROMs that directly incorporate pressure coefficients. Examples include the supremizer approach \cite{Azaiez2024} \cite{Ballarin2015} and Galerkin projection methods in which the momentum equations are projected onto the gradient of a pressure basis $\pbasis$ \cite{Tallet2015} \cite{Leblond2011}.} Recall that the pressure Poisson equation is obtained from the incompressible Navier-Stokes equations by taking the divergence of the momentum equation and invoking the incompressibility constraint.
\begin{equation}
\left\{
\begin{aligned}
\Delta \press
&=
-\nabla\cdot\big((\vit\cdot\nabla)\vit\big)
+\nabla\cdot\mathbf f
&& \text{in } \Omega,
\\
\nabla \press \cdot \mathbf{n}
&=
\left(
\mathbf f
-\partial_t\vit
-(\vit\cdot\nabla)\vit
+\fract{1}{Re}\Delta\vit
\right)\cdot \mathbf n
&& \text{on } \partial\Omega .
\end{aligned}
\right.
\label{eq:ppe}
\end{equation}
Substituting the approximations $\vit_r$ et $\press_r$ into the Poisson equation \eqref{eq:ppe} and performing a Galerkin projection onto the reduced pressure space
$Q_r=\mathrm{span}
\left\{
\pbasis_1,\ldots,\pbasis_{\NbmodeP}
\right\}
$
we obtain
\begin{equation}
\left(
\Delta \press_r,
\pbasis_i
\right)
=-
\left(
\nabla \cdot
\bigl(
(\vit_r\cdot\nabla)\vit_r
\bigr),
\pbasis_i
\right)+
\left(
\nabla\cdot\mathbf{f},
\pbasis_i
\right),
\qquad
i=1,\ldots,\NbmodeP.
\label{eq:ppe_galerkin}
\end{equation}
Substituting the reduced velocity expansion \eqref{eq:velicity_expansion} and reduced pressure expansion \eqref{eq:pressure_expansion} into \eqref{eq:ppe_galerkin} leads to the reduced pressure system
\begin{equation}
\sum_{k=1}^{\NbmodeP}
K_{ik}
\pcoef_k
+
\sum_{j=1}^{\NbmodeU}
\sum_{l=1}^{\NbmodeU}
D_{ijl}
\vcoef_j
\vcoef_l
+
G_i=0,
\qquad
i=1,\ldots,\NbmodeP,
\label{eq:ppe_rom}
\end{equation}
where
\begin{align*}
K_{ik}=
(
\nabla\pbasis_k,
\nabla\pbasis_i
),
\
D_{ijl}
=\left(
\nabla\cdot
\bigl(
(\vbasis_j\cdot\nabla)\vbasis_l
\bigr),
\pbasis_i
\right),
\
G_i
=
-(
\nabla\cdot\mathbf{f},
\pbasis_i
).
\end{align*}
After solving the velocity ROM \eqref{eq:rom_system}, the pressure coefficients are computed from the solution of the pressure ROM \eqref{eq:ppe_rom}.

\section{Non intrusive ROMs based on equations for the Navier–Stokes Problem}
Projection based ROMs rely on the explicit evaluation of the differential operators governing the full order model. The construction of the reduced operators therefore requires access to the discrete representation of the underlying equations, which is not always readily available in practice. Furthermore, the accuracy of the ROM strongly depends on the consistency between the full order and reduced order discretizations. To ensure an accurate projection, the reduced operators should be assembled using the same numerical framework as that employed by the full order solver. Otherwise, inconsistencies between the two discretizations may introduce additional approximation errors and significantly deteriorate the predictive capabilities of the ROM. This requirement is naturally fulfilled in finite element frameworks, where both the full order and ROMs can be derived from the same variational formulation. In contrast, the situation is different for finite volume methods. Since finite volume solvers are not generally formulated within a variational framework, the exact projection of the governing operators onto a reduced basis is often difficult to achieve and may require additional approximations. Another important limitation concerns the use of commercial software. In many industrial codes, the discrete operators and matrices involved in the numerical solution process are not accessible to the user. As a result, the projection step required by intrusive Galerkin-based ROMs cannot be performed, preventing the direct construction of reduced operators. Finally, projection-based approaches fundamentally rely on the availability of the governing equations and their associated operators. Consequently, they cannot be directly applied when only experimental measurements are available. In such situations, where the underlying operators are unknown, intrusive ROM techniques become impractical, motivating the development of non-intrusive and purely data-driven reduced order modeling strategies.

\subsection{Non intrusive ROM for the velocity}
\subsubsection{Methodology\label{ROM_Id_velocityA}}
Let
\[
\mathbf{\vcoef}(t,\param) = \left(\vcoef_1(t,\param), \ldots, \vcoef_{\NbmodeU}(t,\param)\right)^T
\]
denote the vector of reduced velocity coefficients, and
\[
\mathbf{X}
=
\left(
X_1,\ldots,X_{\NbmodeU}
\right)^T,
\]
the vector collecting all coefficients of the ROM. The size of $\mathbf{X}$ is $\NbmodeU^3+\NbmodeU^2+\NbmodeU$ and the expression of $X_i$ is given by :
\[
X_i
=
\left(
\{Q_{ijk}\}_{j,k=1}^{\NbmodeU},
\{L_{ij}\}_{j=1}^{\NbmodeU},
F_i
\right),
\qquad i=1,\ldots,\NbmodeU.
\]
We define the residual function as
\[
\begin{aligned}
\mathcal{R} : \mathbb{R}^{N_X} \times \mathbb{R}^{\NbmodeU} \times \mathbb{R}^{N_{\param}} \times \mathbb{R}^+
&\longrightarrow \mathbb{R}^{\NbmodeU} \\
(\mathbf{X}, \mathbf{\vcoef}, \param,t)
&\longmapsto \mathcal{R}(\mathbf{X}, \mathbf{\vcoef}, \param,t).
\end{aligned}
\]
The \(i\)-th component of \(\mathcal{R}\) is defined by
\[
\mathcal{R}_i
=
\fract{d \vcoef_i}{d t}
+
\sum_{j=1}^{\NbmodeU}\sum_{k=1}^{\NbmodeU}
Q_{ijk}\,\vcoef_j\,\vcoef_k
+
\sum_{j=1}^{\NbmodeU}
L_{ij}\,\vcoef_j
+
F_i,
\qquad i=1,\ldots,\NbmodeU.
\]
With these definitions, the ROM can be written in compact form as
\[
\mathcal{R}\big(\mathbf{X}, \mathbf{\vcoef}(t,\param), \param,t\big) = \mathbf{0}.
\]
The objective of the approach is to identify the coefficients \(\mathbf{X}\) from the values of the reduced coefficients \(\mathbf{\vcoef}\) obtained via POD, which are assumed to be known at \(N_t\) time instants and for \(N_{Re}\) Reynolds number values. These coefficients are denoted by \(\mathbf{\vcoef}^{\mathrm{POD}}(t_n,\param_m)\), for \(n = 1,\ldots,N_t\) and \(m = 1,\ldots,\NbmodeP\).
The identification of the ROM coefficients \(\mathbf{X}\) is formulated as a least-squares minimization problem based on the POD data. More precisely, we seek the vector of ROM coefficients
\(\mathbf{X}\) that minimizes the regularized least-squares functional
\[
\mathbf{X}^\star
=
\underset{\mathbf{X}}{\arg\min}
\;\;
\mathcal{J}(\mathbf{X}),
\]
where
\[
\mathcal{J}(\mathbf{X})
=
\fract{1}{2}
\sum_{k=1}^{\NbmodeP}
\sum_{i=1}^{N_t}
\left\|
\mathcal{R}\big(
\mathbf{X},
\mathbf{\vcoef}^{\mathrm{POD}}(t_i,\param_k),
\param_k,
t_i
\big)
\right\|^2
+
\fract{\lambda}{2}
\left\|
\mathbf{X}
\right\|^2.
\]
The first term measures the discrepancy between the ROM equations and
the POD data, whereas the second term corresponds to a
Tikhonov regularization. The regularization parameter \(\lambda>0\)
controls the trade off between data fitting and solution smoothness.
% This regularization improves the conditioning of the identification
% problem and prevents the estimation of excessively large ROM
% coefficients.
The optimal solution \(\mathbf{X}^\star\) satisfies the first order
optimality condition
\[
\nabla_{\mathbf{X}}\mathcal{J}(\mathbf{X}^\star)=\mathbf{0}.
\]
Since the residual operator is linear with respect to the unknown
coefficient vector \(\mathbf{X}\), the minimization problem can be
recast as a regularized linear least squares problem.
\subsubsection{Regularized linear least squares problem\label{ROM_Id_VelocityB}}
\newcommand{\operOpti}{\mathbf{h}}
The \(i\)-th residual equation can be rewritten as
\[
\mathcal R_i\big(
\mathbf{X},
\mathbf{\vcoef}^{\mathrm{POD}}(t,\param),\param,t\big)
=\fract{{d\vcoef}_i^{\mathrm{POD}}}{dt}(t,\param)
+
\operOpti(t,\param)\,\mathbf{X_i},
\]
where
\[
\operOpti(t,\param)
=
\Big[
\Big\{
\vcoef_j^{\mathrm{POD}}(t,\param)
\vcoef_k^{\mathrm{POD}}(t,\param)
\Big\}_{j,k=1}^{\NbmodeU},
\;
\Big\{
\fract{1}{Re}\vcoef_j^{\mathrm{POD}}(t,\param)
\Big\}_{j=1}^{\NbmodeU},
\;
1
\Big].
\]
Since the POD coefficients are known, the residual is linear with
respect to the unknown vector $\mathbf{X_i}$. Collecting all available
snapshots \((t_n,\param_m)\), \(n=1,\ldots,N_t\),
\(m=1,\ldots,\NbmodeP\), yields the linear system
\begin{equation}
\mathbf{A}\,\mathbf{X_i}=b_i, \qquad i=1,\ldots,\NbmodeU.
\end{equation}
where $A$ is given by
\[
\mathbf{A}=\sum_{n=1}^{N_t}
\sum_{m=1}^{\NbmodeP}\operOpti(t_n,\param_m)
\]
and
\[
b_i
=-
\sum_{n=1}^{N_t}
\sum_{m=1}^{\NbmodeP}\fract{d{\vcoef}_i^{\mathrm{POD}}}{dt}(t_n,\param_m) \\
\]
It is important to note that the vectors $\mathbf{X_i}$ are determined separately, while the matrix $\mathbf{A}$ is common to all equations; only the right-hand side varies. The identification of the ROM coefficients is formulated as
\[
\mathbf{X_i^\star}
=
\underset{\mathbf{X_i}}{\arg\min}
\left\{
\frac12
\left\|
\mathbf{A} \mathbf{X_i}-b_i
\right\|_2^2
+
\fract{\lambda}{2}
\left\|
\mathbf{X_i}
\right\|_2^2
\right\},
\qquad i=1,\ldots,\NbmodeU.
\]
The first order optimality condition yields the normal equations
\[
\left(
\mathbf{A}^{T}\mathbf{A}
+
\lambda \mathbf{I}
\right)
\mathbf{X_i}
=
\mathbf{A}^{T}b_i\qquad i=1,\ldots,\NbmodeU.
\]
The resulting linear systems of size $N_{X_i}\times N_{X_i}$ with $N_{X_i}=\NbmodeU^2+\NbmodeU+1$ are solved iteratively using a
Bi-Conjugate Gradient algorithm.
\subsubsection{Reduced order model\label{ROM_Id_VelocityC}}
The velocity coefficients $\mathbf{\vcoef}$ associated with a new Reynolds number $\param_{new}$, not included in the sampling set used to construct the POD bases, at each time $t$, are then obtained by solving the ROM:
\begin{equation}
\mathcal{R}\big(\mathbf{X^*}, \mathbf{\vcoef}(t,\param_{new}), \param_{new}, t\big) = \mathbf{0}
\label{eq:ROM_IdVelocity}
\end{equation}
where
\[
\mathbf{X^*}
=
\left(
X_1^*,\ldots,X^*_{\NbmodeU}
\right)^T.
\]

\subsection{Non intrusive ROM for the pressure}
\subsubsection{Methodology}
Let
\[
\mathbf{\vcoef}(t,\param) = \left(\vcoef_1(t,\param), \ldots, \vcoef_{\NbmodeU}(t,\param)\right)^T
\]
denote the vector of reduced velocity coefficients,
\[
\mathbf{\pcoef}(t,\param) = \left(\pcoef_1(t,\param), \ldots, \pcoef_{\NbmodeP}(t,\param)\right)^T
\]
the vector reduced pressure coefficients and
\[
\mathbf{X^p}
=
\left(
X^p_1,\ldots,X^p_{\NbmodeP}
\right)^T,
\]
the vector collecting all coefficients of the pressure ROM. The size of $\mathbf{X^P}$ is $N_{X_p}=\NbmodeP^2+\NbmodeP \NbmodeU^2$ and the expression of $X^p_i$ is given by :
\[
X^p_i
=
\left(
\{K_{ij}\}_{j=1}^{\NbmodeP},
\{D_{ijk}\}_{j,k=1}^{\NbmodeU},
G_i
\right),
\qquad i=1,\ldots,\NbmodeP.
\]
We define the residual function as
\[
\begin{aligned}
\mathcal{R}^p : \mathbb{R}^{N_{X_p}} \times \mathbb{R}^{\NbmodeP}\times \mathbb{R}^{\NbmodeU} \times \mathbb{R}^{N_{\param}} \times \mathbb{R}^+
&\longrightarrow \mathbb{R}^{\NbmodeP} \\
(\mathbf{X^p}, \mathbf{\pcoef}, \mathbf{\vcoef}, \param,t)
&\longmapsto \mathcal{R}^p(\mathbf{X_p}, \mathbf{\pcoef}, \mathbf{\vcoef}, \param,t).
\end{aligned}
\]
The \(i\)-th component of \(\mathcal{R}^p\) is defined by
\[
\mathcal{R}_i^p= \sum_{k=1}^{\NbmodeP} K_{ik} \pcoef_k
+ \sum_{j=1}^{\NbmodeU}\sum_{l=1}^{\NbmodeU} D_{ijl}\vcoef_j\vcoef_l+G_i ,
\qquad i=1,\ldots,\NbmodeP.
\]
With these definitions, the ROM can be written in compact form as
\[
\mathcal{R}^p\big(\mathbf{X_p}, \mathbf{\pcoef}, \mathbf{\vcoef}, \param,t\big) = \mathbf{0}.
\]

\subsubsection{Identification of the pressure ROM coefficients \label{ROM_Id_pressure}}
The objective is to identify the ROM coefficient vector $\mathbf{X^p}$ from the reduced velocity and pressure coefficients obtained by Proper Orthogonal Decomposition (POD). These reduced coefficients are assumed to be available at $N_t$ discrete time instants for $N_{\param}$ parameter values. They are respectively denoted by $\mathbf{\vcoef}^{\mathrm{POD}}(t_n,\param_m)$ and $\mathbf{\pcoef}^{\mathrm{POD}}(t_n,\param_m)$, with $n=1,\ldots,N_t$ and $m=1,\ldots,N_{\param}$.

The identification of the ROM coefficients is formulated as a regularized least squares problem. More specifically, the optimal coefficient vector $\mathbf{X^p}$ is defined as the minimizer of the objective function
$$
\mathbf{X^{p\star}}
=
\underset{\mathbf{X^p}}{\arg\min}
\mathcal{J}^p(\mathbf{X^p}),
$$
where
$$
\mathcal{J}^p(\mathbf{X}^{p})
=\fract{1}{2}
\sum_{k=1}^{N_{\param}}
\sum_{i=1}^{N_t}
\left\|
\mathcal{R}^{p}\big(
\mathbf{X}^{p},
\mathbf{\pcoef}^{\mathrm{POD}}(t_i,\param_k),
\mathbf{\vcoef}^{\mathrm{POD}}(t_i,\param_k),
\param_k,
t_i
\big)
\right\|^2
+
\fract{\lambda}{2}
\left\|
\mathbf{X^p}
\right\|^2.
$$
% The regularization parameter ($\lambda>0$) balances the trade-off between data fidelity and solution regularity.
The optimal solution satisfies the first order optimality condition
$$
\nabla_{\mathbf{X^p}}\mathcal{J}^p(\mathbf{X^{p\star}})=\mathbf{0}.
$$
Introducing the Jacobian matrix
$$
\mathbf{J}_{ik}=
\fract{\partial \mathcal{R}^p}
{\partial \mathbf{X^p}}
\Big(
\mathbf{X^p},
\mathbf{\pcoef}^{\mathrm{POD}}(t_i,\param_k),
\mathbf{\vcoef}^{\mathrm{POD}}(t_i,\param_k),
\param_k,
t_i
\Big),
$$
the gradient of the objective function can be expressed as
$$
\nabla_{\mathbf{X^p}}\mathcal{J}^p(\mathbf{X^p})=
\sum_{k=1}^{N_{\param}}
\sum_{i=1}^{N_t}
\mathbf{J}_{ik}^{\top}
\mathcal{R}^p\Big(
\mathbf{X^p},
\mathbf{\pcoef}^{\mathrm{POD}}(t_i,\param_k),
\mathbf{\vcoef}^{\mathrm{POD}}(t_i,\param_k),
\param_k,
t_i
\Big)
+\lambda\mathbf{X^p}.
$$
Since the residual $\mathcal{R}^p$ is linear with respect to the unknown coefficients $\mathbf{X^p}$, the optimality condition reduces to a homogeneous linear system,
$$
\mathbf{A}^p\mathbf{X^{p\star}}=\mathbf{0},
$$
where the matrix $\mathbf{A}^p$ is sparse and rank deficient.
The objective is therefore to determine a non trivial solution belonging to the null space of $\mathbf{A}^p$. Because the homogeneous system admits infinitely many solutions, any vector in $\ker(\mathbf{A}^p)$ satisfies the optimality condition. In practice, the null space is computed by solving the constrained minimization problem
$$
\min_{\mathbf{X^p}\neq\mathbf{0}}
\left\|\mathbf{A}^p\mathbf{X^p}\right \|^2
\quad
\text{subject to}
\quad
\left\|\mathbf{X^p}\right\|=1.
$$
In the present work, this problem is solved by means of the Singular Value Decomposition of $\mathbf{A}^p$,
$$
\mathbf{A}^p=
\mathbf{U}
\mathbf{\Sigma}
\mathbf{V}^{\top}.
$$
The null space is then identified as the subspace spanned by the right singular vectors associated with zero (or numerically negligible) singular values. Consequently, the solution is obtained from the columns of V associated with the smallest singular values.

\subsubsection{Reduced order model}
The pressure coefficients $\mathbf{\pcoef}$ associated with a new Reynolds number $\param_{new}$, not included in the sampling set used to construct the POD bases, at each time $t$, are then obtained by solving the ROM:
\begin{equation}
\mathcal{R}^p\big(\mathbf{X^{p\star}}, \mathbf{\pcoef}(t,\param_{new}), \mathbf{\vcoef}^{ROM}(t,\param_{new}), \param_{new}, t\big) = \mathbf{0}.
\label{eq:ROM_Idpressure}
\end{equation}
where
$\mathbf{\vcoef}^{ROM}=\{\vcoef^{ROM}_1,\ldots,\vcoef^{ROM}_{\NbmodeU}\}$ are the velocity coefficients obtained by solving the velocity ROM \eqref{eq:ROM_IdVelocity}.

\section{Non intrusive ROMs based on equation for porous media problem}
\subsection{Full problem}
For porous media problem the flow is governed by the incompressible Navier-Stokes equations coupled with the Darcy-Forchheimer model to account for momentum losses within the porous medium. Then the associated parametrized non dimensionnal problem consists in finding, the non dimensionnal velocity field $\vit : \Omega \times (0,T) \to \mathbb{R}^d $ and the non dimensionnal pressure $\press : \Omega \times (0,T) \to \mathbb{R}$,
such that
\begin{empheq}[left=\empheqlbrace]{align}
\nabla\cdot\vit
&= 0\\
\frac{\partial \mathbf{u}}{\partial t}
+(\mathbf{u}\cdot\nabla)\mathbf{u}
&=
-\nabla p
+\frac{1}{Re}\Delta\mathbf{u}
-\chi
\left(
\fract{1}{Re\,Da}\mathbf{u}
+
\fract{\tilde{F}}{\sqrt{Da}}
\|\mathbf{u}\|\mathbf{u}
\right)
\end{empheq}
where $Re$ and $Da$ denote the Reynolds and Darcy numbers, respectively,
\begin{equation}
Re=\fract{U_{ref}L_{ref}}{\nu},
\qquad
Da=\fract{\kappa}{L_{ref}^2}.
\end{equation}
$\nu$ is the kinematic viscosity, $\kappa$ the permeability, $U_{ref}$ a reference velocity and $L_{ref}$ a reference length. $\chi$ is a mask function that is equal to zero in the fluid region, and is equal to one in the porous region. When $\chi=1$, two terms are added in the momentum equation of the porous region, the Darcy term and the Forchheimer term. The Darcy term takes into account the viscous resistance and depends on the dynamic viscosity $\nu$ and the permeability $\kappa$. The Forchheimer term represents the inertial resistance, it depends on the inertial factor which reads:
\begin{equation}
\tilde{F} = \frac{1.75}{\displaystyle{\sqrt{150}}}\epsilon^{-3/2}.
\end{equation}
where $\epsilon$ is the porosity.
\subsection{Non intrusive ROMs}
\subsubsection{Generalities}
It is important to note that the projection based Galerkin ROM approach is not well suited for this class of problems. Indeed, the Forchheimer term cannot be directly expressed in reduced form during the Galerkin projection stage. At each time step, the norm $\left\|\mathbf{u}\right\|$ must be recomputed and projected, which significantly increases the computational cost and renders the ROM inefficient. To overcome this difficulty, the projection of this term onto the reduced basis can be handled using the Discrete Empirical Interpolation Method  \cite{German2021}, or alternatively, these projected terms can be approximated using artificial neural networks , where the inputs are the coefficients of the ROM \cite{allery_beghein_2025}. The ROM construction approach by identification proposed in this work allows us to circumvent this issue.
\subsubsection{Methodology}
We recall that
\[
\mathbf{\vcoef}(t,\param,Da) = \left(\vcoef_1(t,\param), \ldots, \vcoef_{\NbmodeU}(t,\param,Da)\right)^T
\]
denotes the vector of reduced velocity coefficients. For the porous media problem, the vector collecting all ROM coefficients,
\[
\mathbf{X}
=
\left(
X_1,\ldots,X_{\NbmodeU}
\right)^T,
\]
has dimension $2\NbmodeU^3 + 2\NbmodeU^2 + \NbmodeU$, and each component $X_i$ is defined as
\[
X_i
=
\left(
\{Q_{ijk}\}_{j,k=1}^{\NbmodeU},\{Q^{F}_{ijk}\}_{j,k=1}^{\NbmodeU},
\{L_{ij}\}_{j=1}^{\NbmodeU},\{L^D_{ij}\}_{j=1}^{\NbmodeU},
F_i
\right),
\qquad i=1,\ldots,\NbmodeU.
\]
In this case, the residual function is defined as
\[
\begin{aligned}
\mathcal{R} : \mathbb{R}^{N_X} \times \mathbb{R}^{\NbmodeU} \times \mathbb{R}^{N_{\param}} \times \mathbb{R}^{N_{Da}} \times \mathbb{R}^+
&\longrightarrow \mathbb{R}^{\NbmodeU} \\
(\mathbf{X}, \mathbf{\vcoef}, \param, Da, t)
&\longmapsto \mathcal{R}(\mathbf{X}, \mathbf{\vcoef}, \param, Da, t).
\end{aligned}
\]
The \(i\)-th component of \(\mathcal{R}\) is given by
\[
\mathcal{R}_i
=
\fract{d \vcoef_i}{d t}
+
\sum_{j=1}^{\NbmodeU}\sum_{k=1}^{\NbmodeU}
\big(Q_{ijk}+\fract{1}{Da}Q^{F}_{ijk}\big)\,\vcoef_j\,\vcoef_k
+
\sum_{j=1}^{\NbmodeU}
\big(\fract{1}{Re}L_{ij}+\fract{1}{ReDa}L_{ij}^D\big)\,\vcoef_j
+
F_i,
\qquad i=1,\ldots,\NbmodeU.
\]
With these definitions, the ROM can be written\footnote{The methodology has been presented here with two varying parameters, $\param$ and $Da$. However, in the numerical results section, only one parameter is varied, namely the Reynolds number, while the Darcy number is kept fixed. The two-parameter study will be the subject of future work.} in compact form as
\[
\mathcal{R}\big(\mathbf{X}, \mathbf{\vcoef}(t,\param,Da), \param, Da, t\big) = \mathbf{0}.
\]
The coefficients $\mathbf{X}$ are identified from the velocity reduced coefficients $\mathbf{\vcoef}$ obtained via POD, assumed to be known at $N_t$ time instants, for $N_{Re}$ Reynolds number values and $N_{Da}$ Darcy number values, using the same methodology as described in Sections \ref{ROM_Id_VelocityB} and \ref{ROM_Id_VelocityC}.

~\\
The methodology for determining the reduced pressure coefficients is identical to that used for the Navier--Stokes problem (see Section \ref{ROM_Id_pressure}).

\section{Results}
\subsection{Results obtained for the lid driven cavity case}
\subsubsection{Studied configuration}
\begin{figure}[!h]
	\centering		
	\includegraphics[scale=0.3]{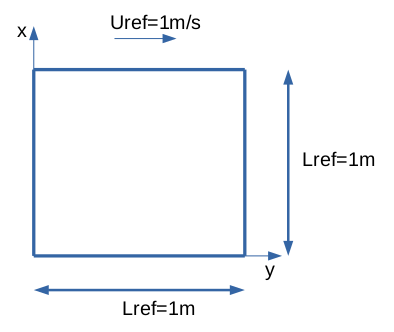} 
	\caption{Lid driven cavity}	
	\label{cavity}	
\end{figure}

In this section, a two-dimensional square cavity of side $L_{ref}=1m$ is considered (see Figure \ref{cavity}). The top wall moves at a constant velocity $U_{ref}=1m/s$, while the other walls are stationary. Each case studied is characterized by the Reynolds number $Re=\frac{U_{ref} L_{ref}}{\nu}$, where $\nu$ is the kinematic viscosity. Several simulations with various Reynolds numbers have been considered : $Re \in \{8750, 9000, 9250, 9500, 9750\}$. The flow is unsteady, laminar and incompressible. Computations have been performed with the unsteady incompressible solver pisoFoam of openFOAM, on a non-uniform grid of $300\times 300$ cells. The computations have been validated for a Reynolds number of 10000. For this configuration, the horizontal velocity at $x=0.125,y=0.8125$ exhibits a periodic behavior, and the period found with our computations is $1.675s$, which is close to the period of $1.64s$ obtained by Bruneau and Saad \cite{bruneau_2006}.

In the following, to quantify the discrepancy between the reference solution\footnote{$h$ can be the velocity $\vit$ or the pressure $p$.}
\(h(x,t)\) and the predicted solution \(\hat{h}(x,t)\),
we consider the absolute spatial \(L^2\)-error at time \(t\), defined by
$$e_{L^2}(t)=\left\|h(\cdot,t)-\hat{h}(\cdot,t)\right\|_{L^2(\Omega)},$$
and the relative spatial \(L^2\)-error, defined by
$$\varepsilon_{L^2}(t)=\frac{
\left\|h(\cdot,t)-\hat{h}(\cdot,t)\right\|_{L^2(\Omega)}
}{
\left\|h(\cdot,t)\right\|_{L^2(\Omega)}
}.$$
Here, the \(L^2\)-norm over the spatial domain \(\Omega\) is defined as
$$
\left\|h(\cdot,t)\right\|_{L^2(\Omega)}
=
\left(
\int_{\Omega}
\left|h(x,t)\right|^2\,\mathrm{d}x
\right)^{1/2}.
$$
Given \(N_t\) time instants
\(\{t_i\}_{i=1}^{N_t}\), the mean absolute \(L^2\)-error is defined as
\begin{equation}
\overline{e}_{L^2}
=\frac{1}{N_t}
\sum_{i=1}^{N_t}
e_{L^2}(t_i),
\label{eq:mean_absolute_L2_error}
\end{equation}
while the mean relative \(L^2\)-error is given by
\begin{equation}
\overline{\varepsilon}_{L^2}=\frac{1}{N_t}
\sum_{i=1}^{N_t}
\varepsilon_{L^2}(t_i).
\label{eq:mean_relative_L2_error}
\end{equation}
The maximum absolute \(L^2\)-error over the considered time instants is defined by
\begin{equation}
e_{L^2}^{\max}=\max_{1\leq i\leq N_t}
e_{L^2}(t_i),
\label{eq:max_absolute_L2_error}
\end{equation}
and the maximum relative \(L^2\)-error is defined by
\begin{equation}
\varepsilon_{L^2}^{\max}=\max_{1\leq i\leq N_t}
\varepsilon_{L^2}(t_i).
\label{eq:max_relative_L2_error}
\end{equation}
.
\subsubsection{Results obtained for a fixed Reynolds number}
This paragraph deals with the prediction by the reduced order model of the velocity and pressure fields for a Reynolds number of 9500. 
The velocity and pressure POD bases are generated from 103 snapshots uniformly sampled over a time period of $T = 10.3\,\mathrm{s}$, with a sampling interval of $\Delta t = 0.1\,\mathrm{s}$. The ROM formulation requires the evaluation of the temporal derivatives of the velocity POD coefficients, $\alpha_{\mathrm{POD}}^i$. Since these derivatives are computed \emph{a posteriori}\footnote{A more accurate alternative would consist in projecting the time derivative of the velocity field, available during the full-order simulation in OpenFOAM, onto the velocity POD basis.}, their accuracy depends on the snapshot sampling interval. To improve the estimation of the temporal derivatives, the POD coefficient dataset is temporally enriched by interpolation. In this study, the derivatives are approximated using a first-order finite-difference scheme with a time step of $\Delta t = 10^{-5}\,\mathrm{s}$. Finally, the ROM is trained using only $50\%$ of the coefficients from the enriched dataset, randomly sampled over the interval $[0,\,5\,\mathrm{s}]$.\\

\begin{figure}[!htbp]
	\centering
	
	\begin{subfigure}{0.43\textwidth}
		\centering
		\includegraphics[width=\textwidth]{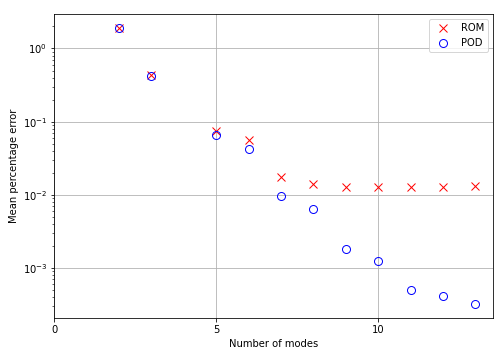}
		\caption{\small Reconstruction error}
	\end{subfigure}
	\hfill
	\begin{subfigure}{0.5\textwidth}
		\centering
		\includegraphics[width=\textwidth]{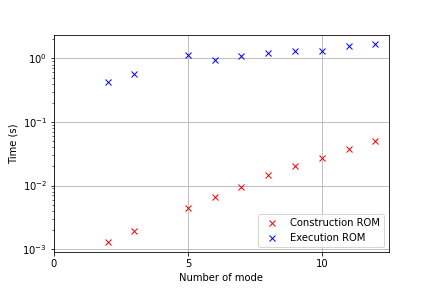}
		\caption{\small Computing time }
	\end{subfigure}
	\caption{Left: mean percentage velocity reconstruction error as a function of the number of modes for the exact POD reconstruction and the ROM. Right: computational times for ROM construction and simulation ($Re=9500$).}
	\label{fig:lid_cavity_unparam_vitesse}
\end{figure}

\paragraph{a) ROM for velocity}~\\
\indent In Figure \ref{fig:lid_cavity_unparam_vitesse} (left), the mean relative velocity reconstruction
error is plotted as a function of the number of retained POD modes, both for the exact POD reconstruction and for the ROM prediction. The error is evaluated over the entire POD sampling time interval, $[0,10.3]$, while the ROM is constructed using only the time interval $[0,5]$.

The reconstruction error obtained with the ROM is almost identical to that obtained with the exact POD reconstruction up to 7 modes, corresponding to a mean relative error of approximately $0.01\%$.
Beyond 7 modes, the error remains almost unchanged. These results show that the identification approach enables the construction of a robust and accurate velocity ROM, even when it is evaluated beyond the time interval used for its construction.

Figure \ref{fig:lid_cavity_unparam_vitesse} (right) shows the computational time required for model construction and simulation as a function of the number of modes. As expected, both computational times increase with the number of modes. However, they remain very low. For 10 modes, the model construction takes approximately $0.02$ s, while the simulation takes only $0.4$ s. The proposed approach is therefore very fast, with both the construction and simulation of the ROM being performed in a quasi-instantaneous manner. In particular, the construction time remains negligible compared with a Galerkin projection approach, which requires the computation of the different differential operators.

\begin{figure}[!htbp]
	\centering
	\begin{subfigure}{0.48\textwidth}
		\centering
		\includegraphics[width=\textwidth]{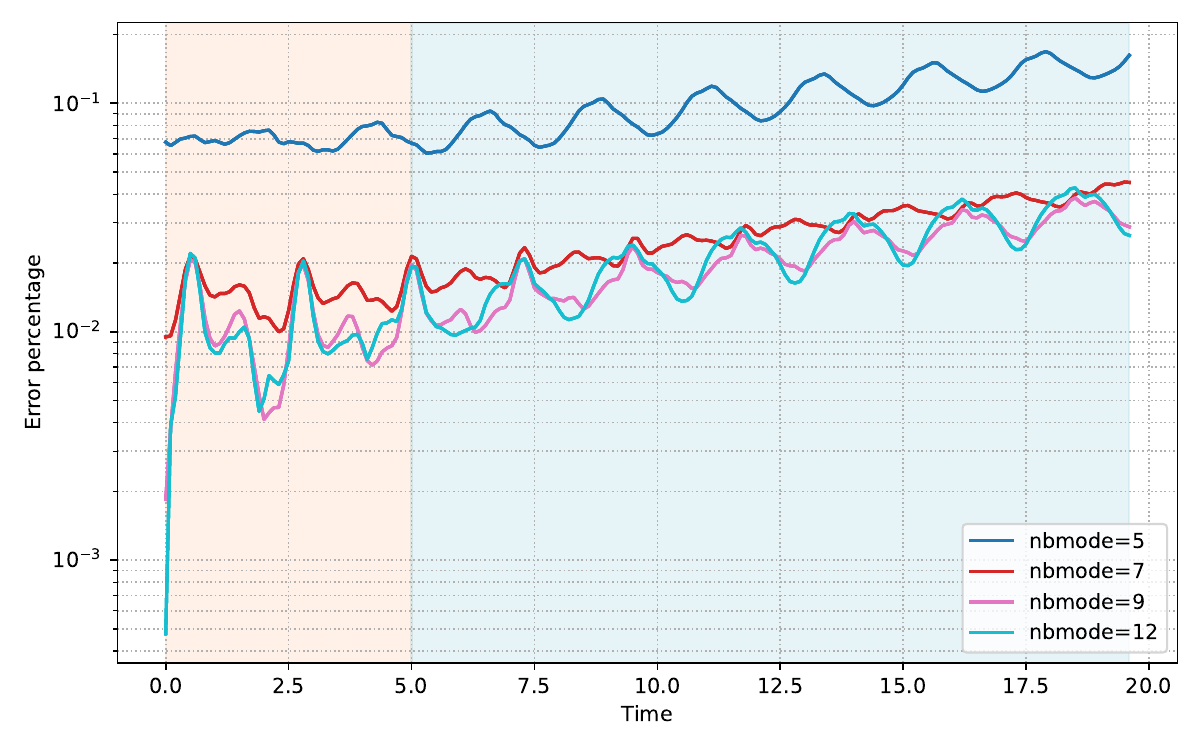}
		\caption{\small Reconstruction error}
	\end{subfigure}
	\hfill
	\begin{subfigure}{0.48\textwidth}
		\centering
		\includegraphics[width=\textwidth]{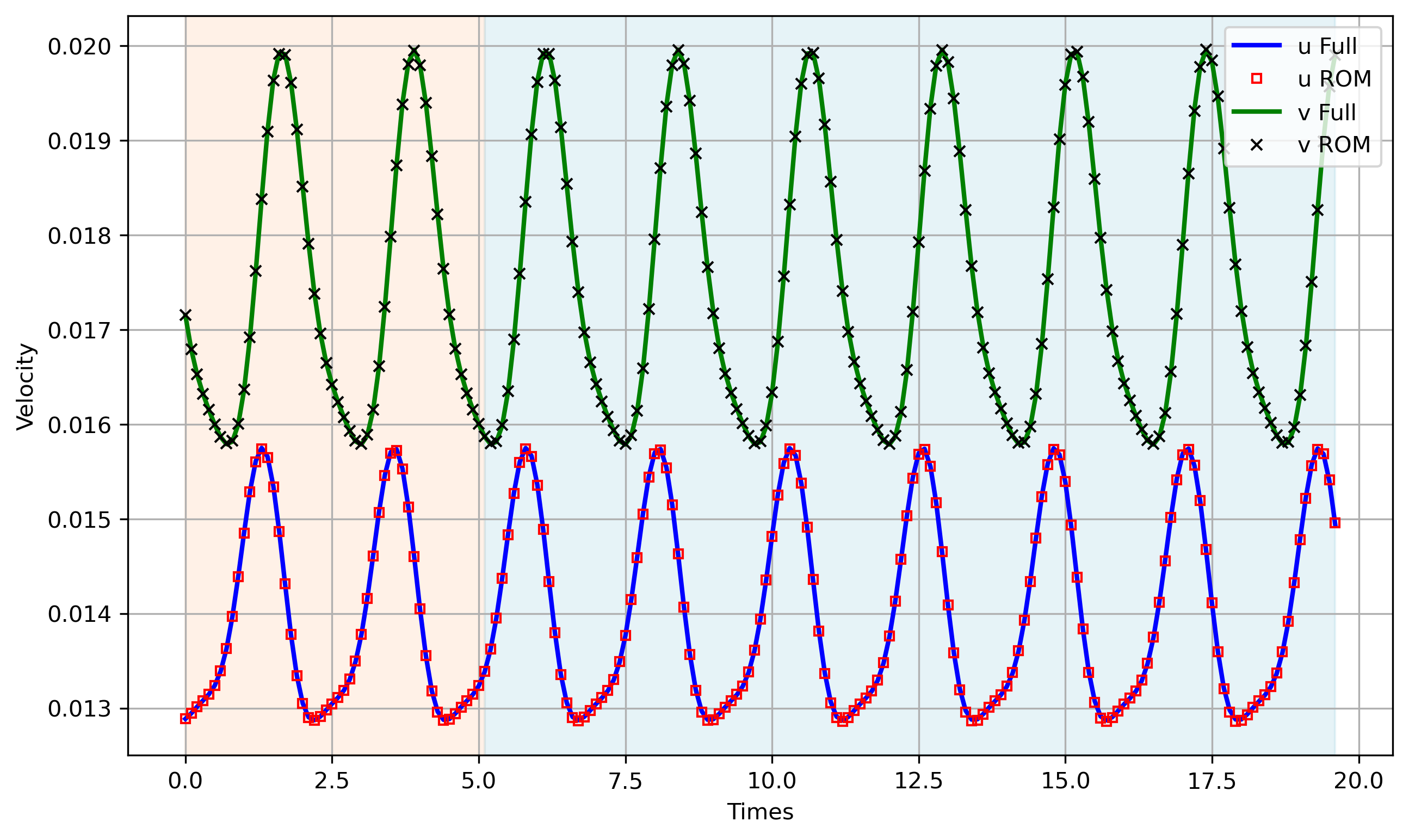}
		\caption{\small Probe 1 }
	\end{subfigure}
	
	\vspace{0.5cm}
	
	\begin{subfigure}{0.48\textwidth}
		\centering
		\includegraphics[width=\textwidth]{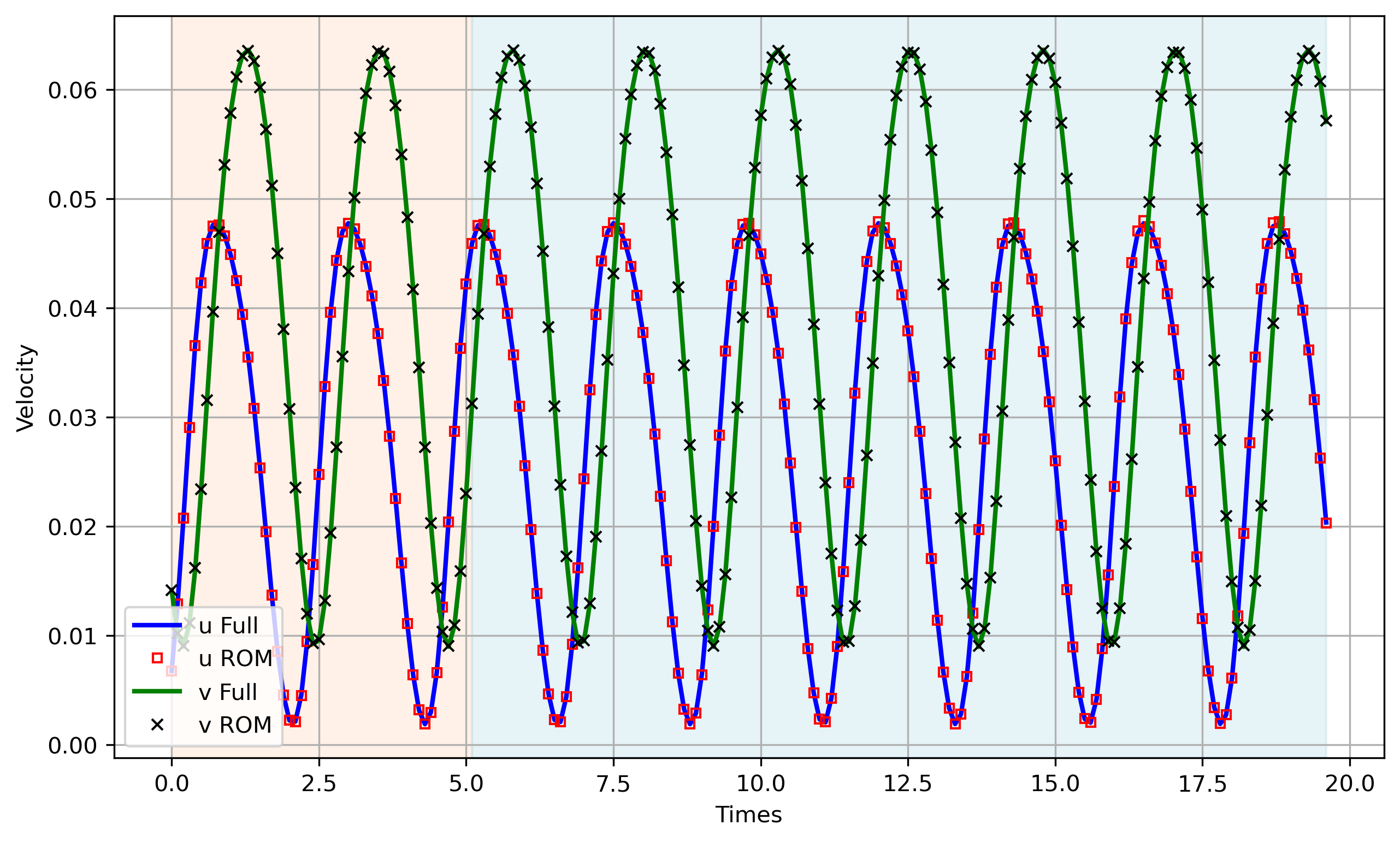}
		\caption{\small Probe 2 }
	\end{subfigure}
	\hfill
	\begin{subfigure}{0.48\textwidth}
		\centering
		\includegraphics[width=\textwidth]{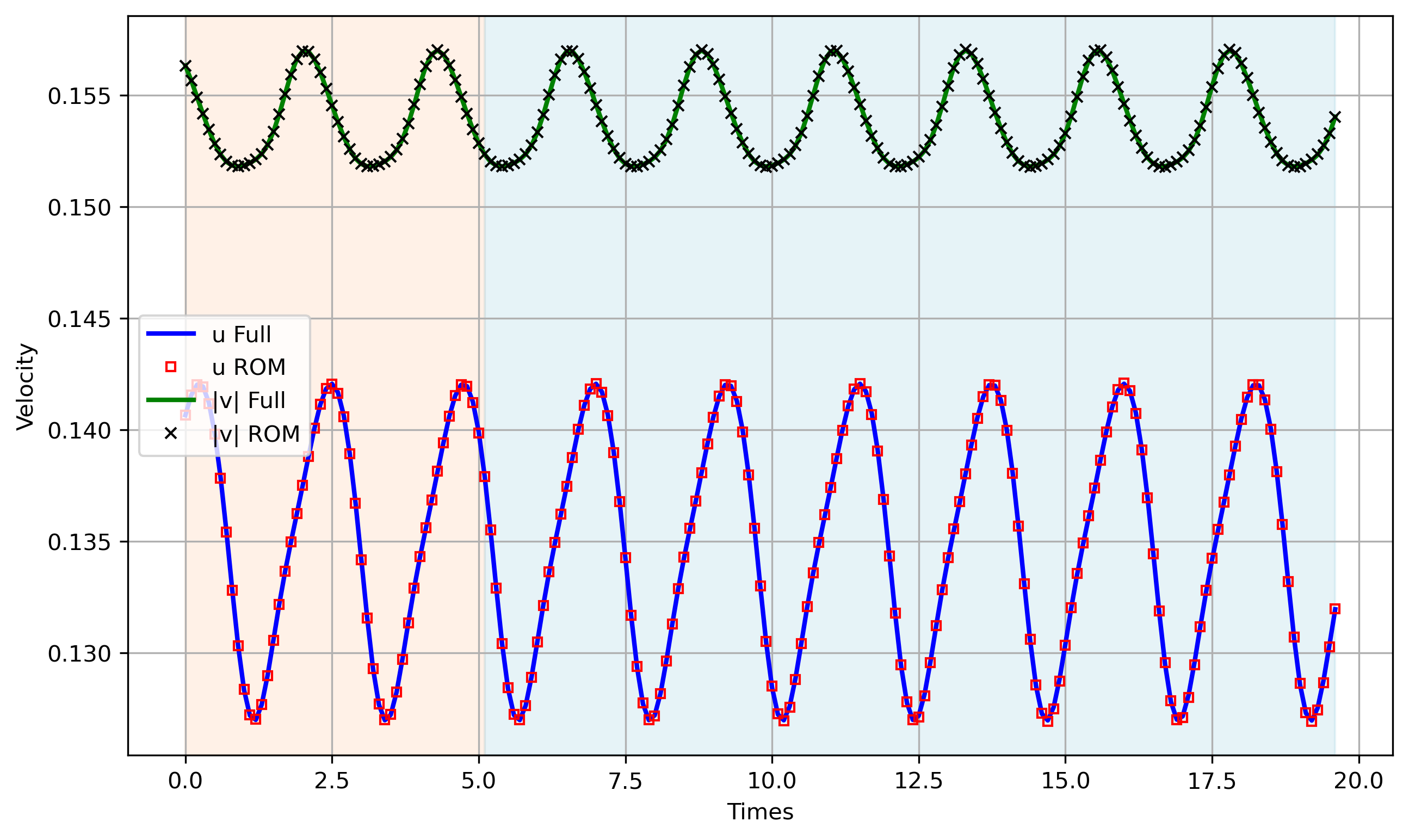}
		\caption{\small Probe 3 }
	\end{subfigure}
	
	\caption{Reconstruction error as a function of the number of modes and the velocity components (horizontal u and vertical v) measured at the probe locations for $Re=9500$ and $\NbmodeU=9$. Light orange background: ROM training period; light blue background: ROM prediction period.}
	\label{fig:lid_cavity_unparam}
\end{figure}

\begin{figure}[!htbp]
	\centering
	
	\begin{subfigure}{0.48\textwidth}
		\centering
		\includegraphics[width=\textwidth]{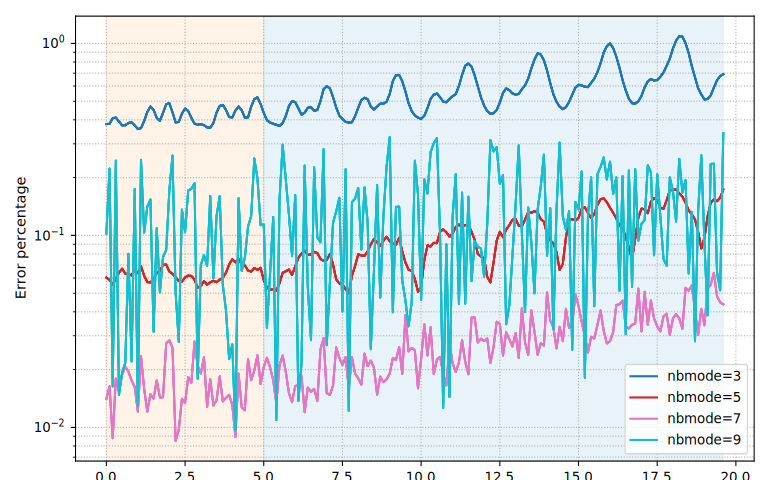}
		\caption{\small Reconstruction error}
	\end{subfigure}
	\hfill
	\begin{subfigure}{0.48\textwidth}
		\centering
		\includegraphics[width=\textwidth]{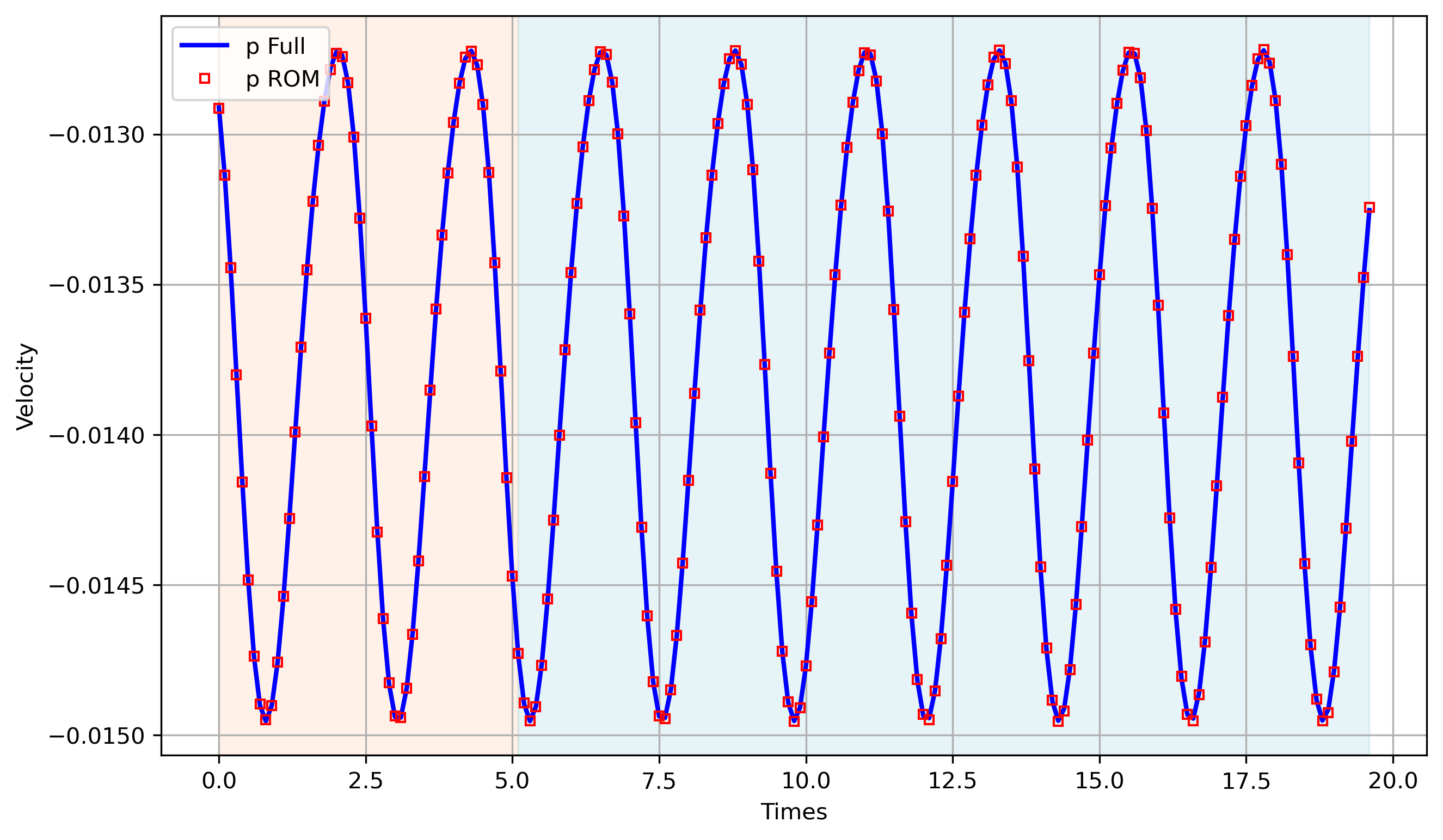}
		\caption{\small Probe 1 }
	\end{subfigure}
	
	\vspace{0.5cm}
	
	\begin{subfigure}{0.48\textwidth}
		\centering
		\includegraphics[width=\textwidth]{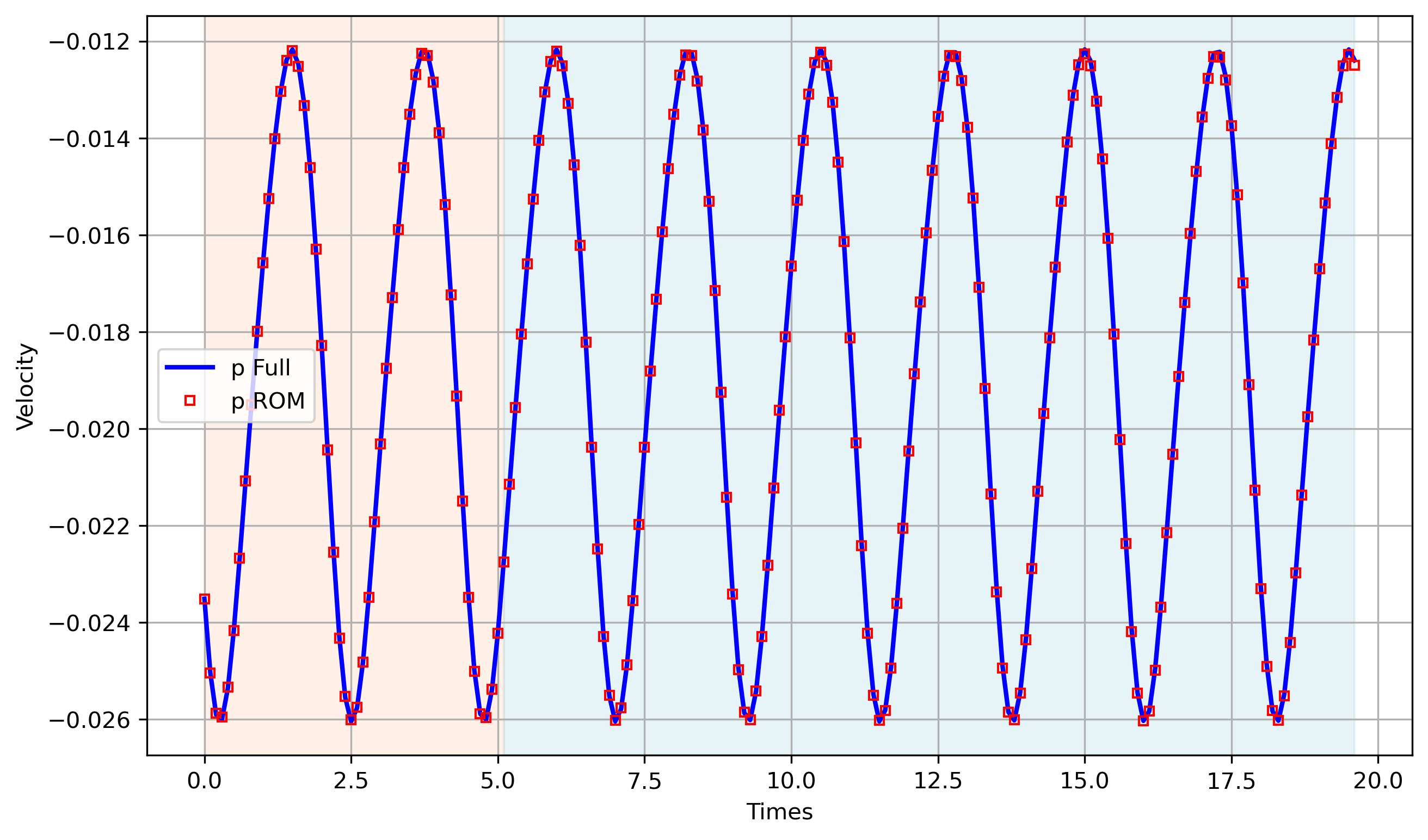}
		\caption{\small Probe 2  }
	\end{subfigure}
	\hfill
	\begin{subfigure}{0.48\textwidth}
		\centering
		\includegraphics[width=\textwidth]{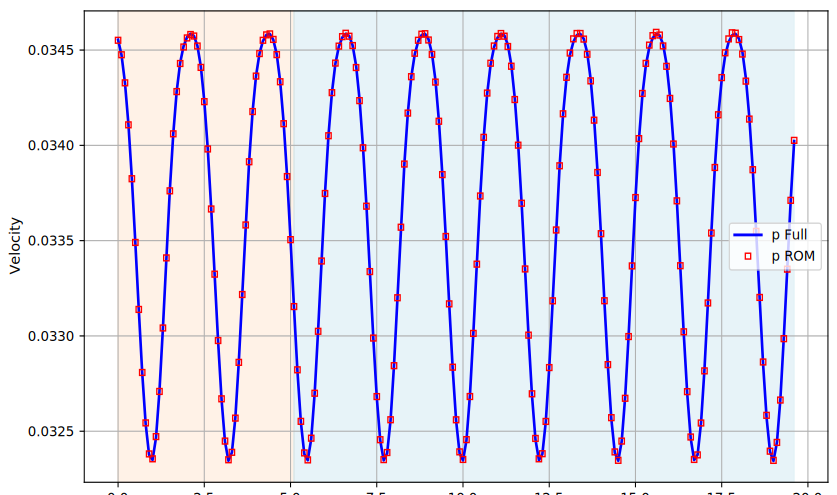}
		\caption{\small Probe 3  }
	\end{subfigure}
	
	\caption{Reconstruction error as a function of the number of modes and pressure measured at the probe locations for $Re=9500$ and $N_u=N_p=7$. Light orange background: ROM training period; light blue background: ROM prediction period.}
	\label{fig:lid_cavity_unparam_pressure}
\end{figure}

To assess the robustness of the ROM, we track the temporal evolution of the velocity components (horizontal $u$ and vertical $v$)  and the pressure $p$ at three probes located near the corners of the cavity, where the flow dynamics are most complex and time-dependent. The three probes are defined as follows:
\begin{itemize}
	\item[-]  probe 1: lower-right corner, $(x,y)= (0.9, 0.1)$
	\item[-]  probe 2: upper-left corner, $(x,y)= (0.1, 0.9)$
	\item[-]  probe 3: upper-right corner, $(x,y)= (0.9, 0.9)$
\end{itemize}
In figure \ref{fig:lid_cavity_unparam}, the time evolutions of the reconstruction error (defined by Equ. \ref{eq:mean_relative_L2_error}), and of the velocity at probe 1, probe 2, and probe 3 are displayed. A time interval of $5$ s is used for the training period, while a time interval of $19.8$ s is considered for the prediction period. The reconstruction error increases only slightly with time, indicating that the ROM remains stable over the prediction interval. As expected, the error is higher when a smaller number of POD modes is used. For 7 modes, the error at the end of the simulation remains below $0.03\%$. The predicted horizontal and vertical velocity components obtained with 9 modes are in very good agreement with the reference values at the three probe locations.
\paragraph{b) ROM for pressure}~\\
\indent
Figure \ref{fig:lid_cavity_unparam_pressure} compares the mean relative pressure reconstruction error obtained with the exact POD representation and with two pressure ROM configurations as the number of retained pressure POD modes is varied. In the first configuration, denoted as ROM, the number of velocity modes used to compute the right-hand side of the pressure ROM is chosen equal to the number of pressure modes retained for its construction. In the second configuration, the number of velocity modes is fixed to 7. The comparison is performed over the complete POD sampling interval, $[0,10.3]$, although the pressure ROM is identified using data only from the initial interval $[0,5]$.

The ROM using 7 velocity modes (green curve) provides a reconstruction error close to that of the exact POD representation up to 7 pressure modes, with a mean relative error of approximately $0.02\%$, beyond which the error remains nearly unchanged. These results indicate that fixing the number of velocity modes to 7 provides a stable and accurate pressure reconstruction. In contrast, when the number of velocity modes is varied together with the number of pressure modes (red curve), the reconstruction error exhibits a less regular behavior. Up to 7 modes, the error remains generally close to that of the POD reconstruction, except for 6 modes, where a larger error is observed. Beyond 7 modes, the error increases with the number of modes. This behavior is explained by the dependence of the pressure ROM on the velocity prediction.
\begin{figure}[!htbp]
	\centering
	\includegraphics[scale=0.4]{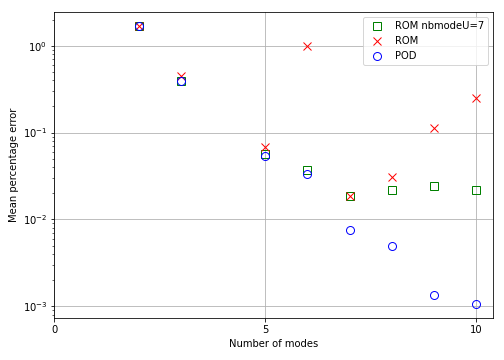}
	\caption{Mean percentage pressure reconstruction error as a function of the number of modes for the exact POD reconstruction and the ROM.}
	\label{fig:lid_cavity_unparam_pression}
\end{figure}

Figure \ref{fig:lid_cavity_unparam_pressure} presents the temporal evolution of the reconstruction error and the reconstructed pressure at probes 1, 2, and 3. As observed for the velocity reconstruction, the pressure error increases only slightly over time. With 7 modes, the predicted pressure remains in very good agreement with the reference solution at all three probe locations.

\subsubsection{Results obtained by the parametric ROM}
These paragraphs investigate the numerical prediction of velocity and pressure with the reduced order models for parametric simulations, where the parameter is the Reynolds number. To assess the accuracy and efficiency of the ROM, five cases are considered:
\begin{itemize}
\item[-] Case A: training set $Re \in \{9000, 9250, 9500, 9750\}$, with $Re_{\text{test}} = 8750$.
\item[-] Case B: training set $Re \in \{8750, 9250, 9500, 9750\}$, with $Re_{\text{test}} = 9000$.
\item[-] Case C: training set $Re \in \{8750, 9000, 9500, 9750\}$, with $Re_{\text{test}} = 9250$.
\item[-] Case D: training set $Re \in \{8750, 9000, 9250, 9750\}$, with $Re_{\text{test}} = 9500$.
\item[-] Case E: training set $Re \in \{8750, 9000, 9250, 9500\}$, with $Re_{\text{test}} = 9750$.
\end{itemize}
It is worth noting that Cases B–D correspond to interpolation settings, whereas Cases A and E correspond to extrapolation settings, since the test Reynolds numbers lie outside the range of the training set.

The velocity and pressure POD bases were constructed using 103 snapshots for each Reynolds number, uniformly spaced at intervals of $0.1 s$.

\paragraph{a) ROM for the velocity}~\\

For each case, the ROM associated with the momentum equation is constructed using 40 velocity POD modes associated with the four Reynolds numbers contained in the training set. To improve the estimation of the temporal derivatives of the velocity POD coefficients, $\alpha_{\mathrm{POD}}^i$, involved in the ROM formulation, the POD coefficient dataset is enriched through temporal interpolation using a non-dimensional time step of $\Delta t^{*}=0.0001$. For each Reynolds number, only $70\%$ of the coefficients from the enriched dataset, randomly selected over the interval $[0,\,10.3]$, are retained for the ROM construction. The POD coefficients $a_i^{\mathrm{POD}}(t_j, Re_k)$ are therefore considered for $1 \leq i \leq 40$, $1 \leq k \leq 4$, and $1 \leq j \leq N_{\mathrm{build}}$, where $N_{\mathrm{build}}$ denotes the number of retained time instances after random subsampling of the full POD temporal dataset.

In table \ref{tab:rec_error_all}, the mean absolute error and the maximum error for the velocity reconstruction are reported. As expected, Cases A and E, which correspond to extrapolation settings, exhibit larger errors than Cases B--D, which correspond to interpolation settings. Nevertheless, the errors remain low for all cases. For the interpolation cases, the mean relative errors range from $0.126\%$ to $0.222\%$, while the maximum relative errors remain below $0.71\%$. For the extrapolation cases, the mean relative errors remain below $0.84\%$, with maximum relative errors of $2.27\%$ and $2.72\%$ for Cases A and E, respectively. Thus, the mean relative error remains below $1\%$ for all cases, and the maximum relative error remains below $3\%$. These results show that the ROM provides an accurate reconstruction of the velocity field over the range of tested Reynolds numbers. Although extrapolation leads to higher errors than interpolation, the increase remains limited.
\begin{table}[!htbp]
	\centering
	\begin{tabular}{cccccc}
		\toprule
		Training& Test Re & Mean abs. & Max abs. & Mean $\%$ & Max $\%$ \\
		\midrule
		\rowcolor{lightgray}
		Case A & 8750 & $2.15 \times 10^{-4}$ & $6.97 \times 10^{-4}$ & $6.98 \times 10^{-1}$ & $2.27$ \\
		Case B & 9000 & $4.95 \times 10^{-5}$ & $1.78 \times 10^{-4}$ & $1.61 \times 10^{-1}$ & $5.79 \times 10^{-1}$ \\
		\rowcolor{lightgray}
		Case C & 9250 & $3.85 \times 10^{-5}$ & $1.33 \times 10^{-4}$ & $1.26 \times 10^{-1}$ & $4.34 \times 10^{-1}$ \\
		Case D & 9500 & $6.80 \times 10^{-5}$ & $2.16 \times 10^{-4}$ & $2.22 \times 10^{-1}$ & $7.06 \times 10^{-1}$ \\
		\rowcolor{lightgray}
		Case E &
		9750 & $2.55 \times 10^{-4}$ & $8.31 \times 10^{-4}$ & $8.38 \times 10^{-1}$ & $2.72$ \\
		\bottomrule
	\end{tabular}
	\caption{Velocity reconstruction errors (mean absolute error, max absolute error, mean percentage error, and max percentage error) for different sampling cases used to construct the ROM. Cases A and E correspond to extrapolation settings, whereas Cases B--D correspond to interpolation settings.}
	\label{tab:rec_error_all}
\end{table}

\begin{figure}[!htbp]
	\centering
	\begin{subfigure}{0.48\textwidth}
		\centering
		\includegraphics[width=\textwidth]{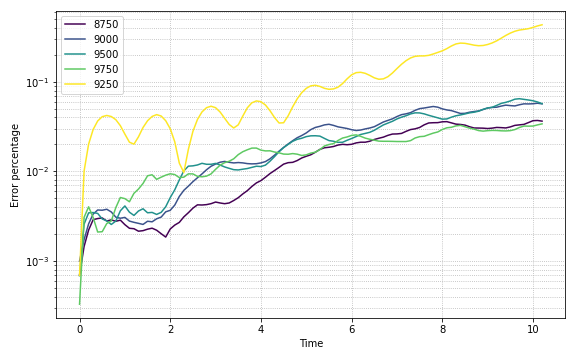}
		\caption{\small Reconstruction error}
	\end{subfigure}
	\hfill
	\begin{subfigure}{0.48\textwidth}
		\centering
		\includegraphics[width=\textwidth]{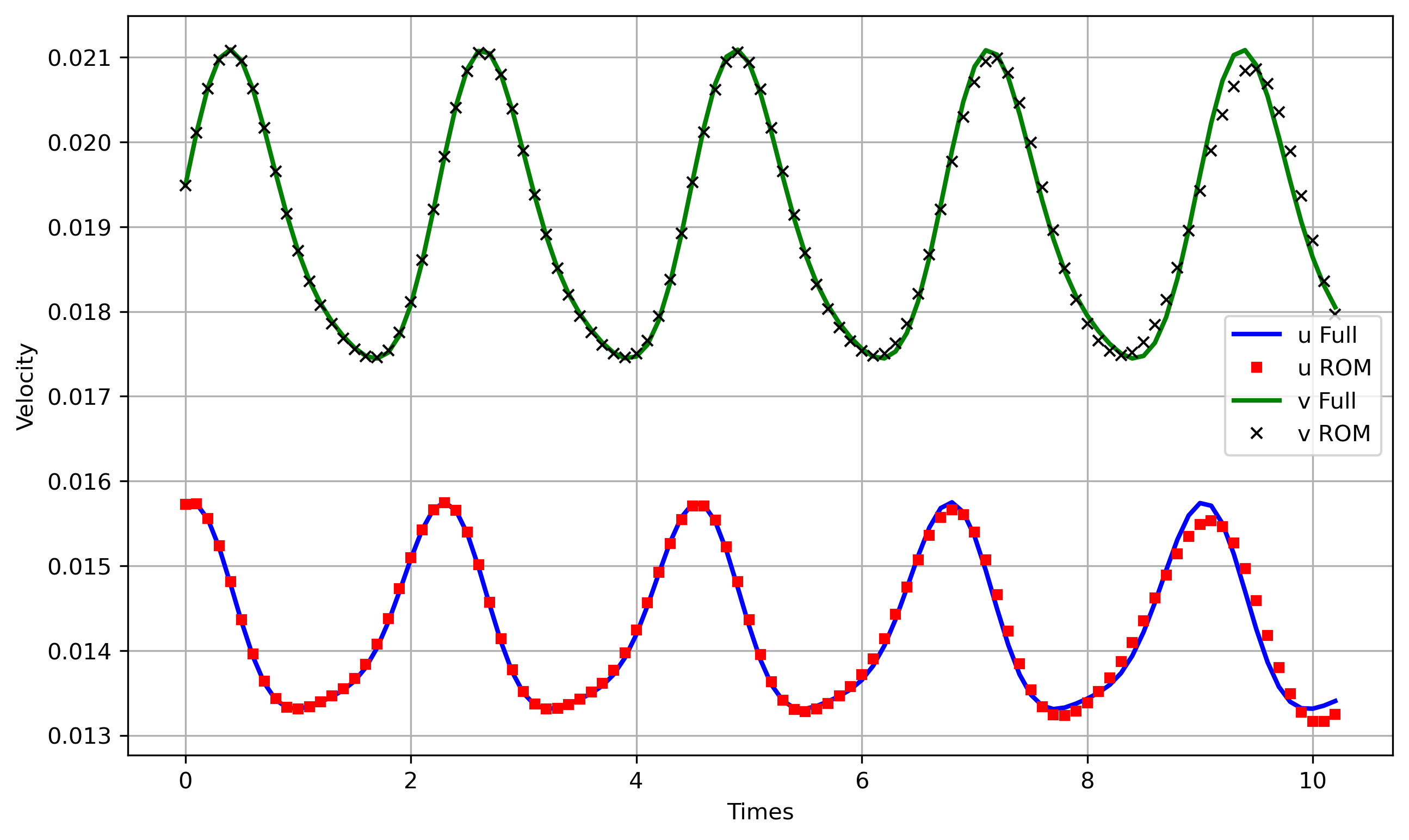}
		\caption{\small Probe 1 -- $Re=9250$}
	\end{subfigure}
	
	\vspace{0.5cm}
	
	\begin{subfigure}{0.48\textwidth}
		\centering
		\includegraphics[width=\textwidth]{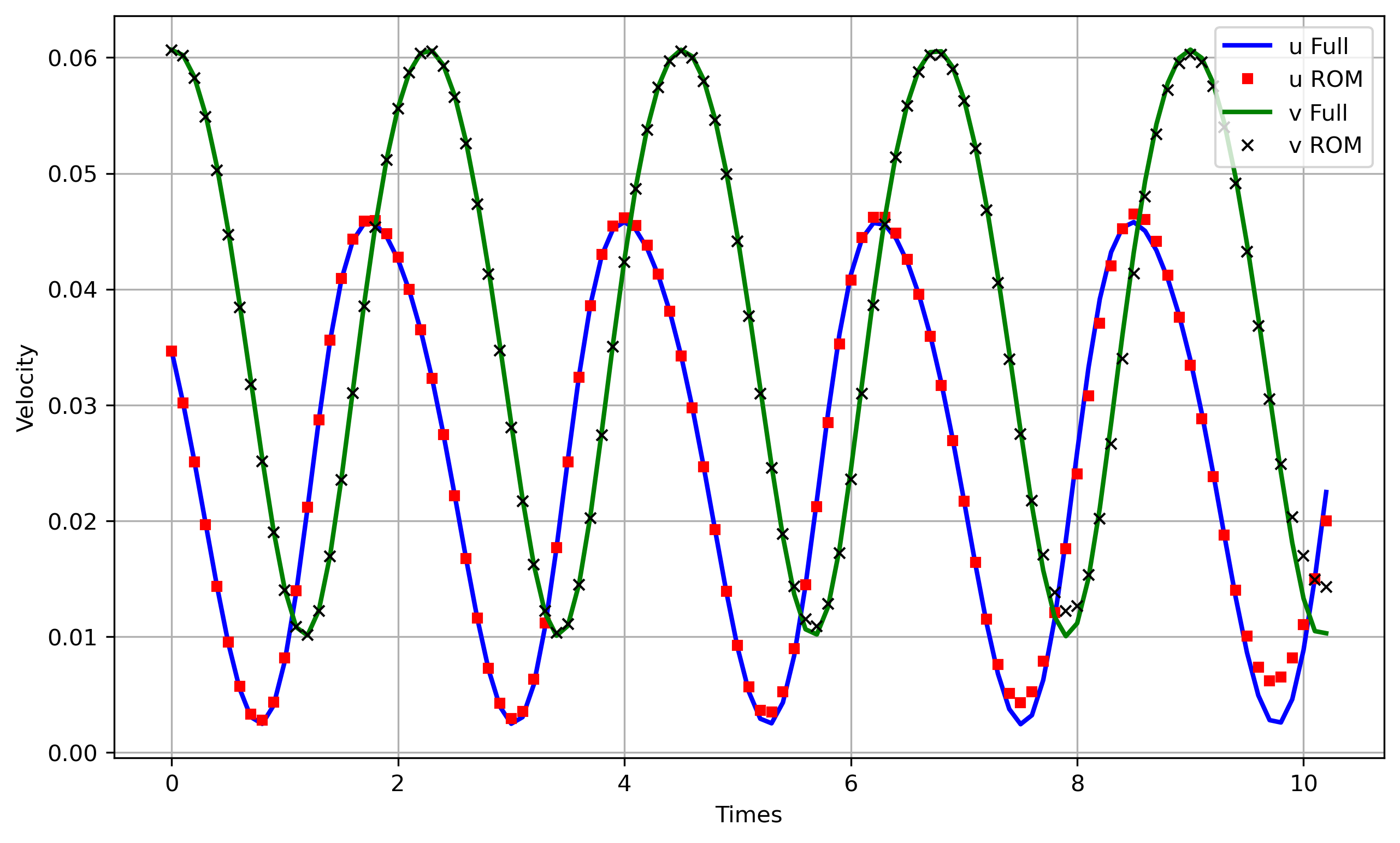}
		\caption{\small Probe 2 -- $Re=9250$ }
	\end{subfigure}
	\hfill
	\begin{subfigure}{0.48\textwidth}
		\centering
		\includegraphics[width=\textwidth]{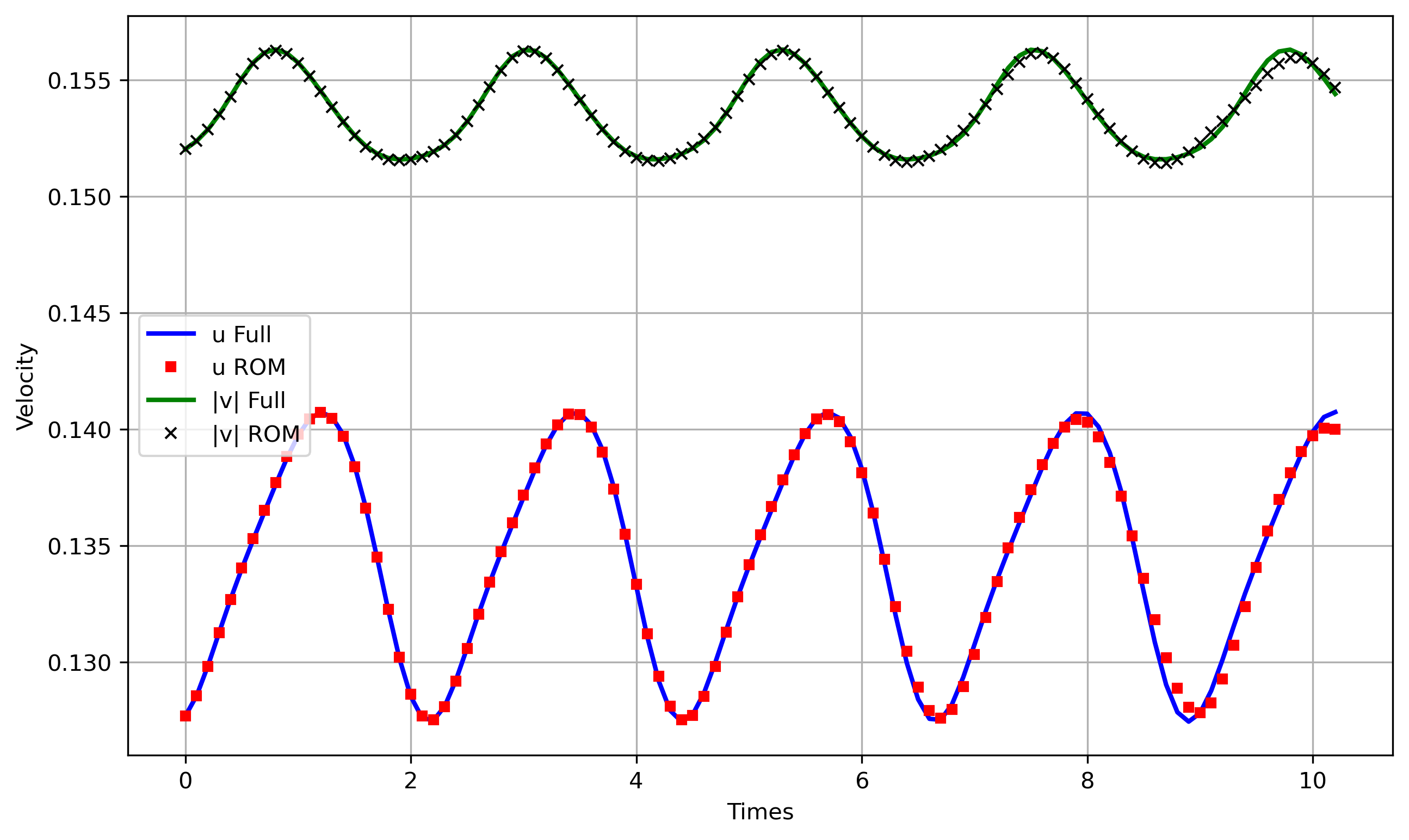}
		\caption{\small Probe 3 -- $Re=9250$ }
	\end{subfigure}
	
	\caption{Reconstruction error and velocity components (horizontal $u$, vertical $v$) measured at probes for Case C (sampling $Re \in \{8750, 9000, 9500, 9750\}$).}
	\label{fig:lid_cavity_caseC}
\end{figure}

\begin{figure}[!htbp]
	\centering
	
	\begin{subfigure}{0.48\textwidth}
		\centering
		\includegraphics[width=\textwidth]{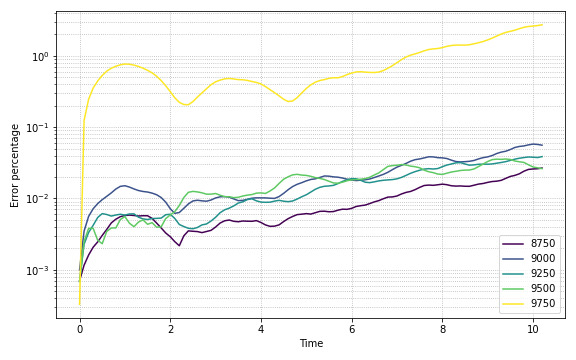}
		\caption{\small Reconstruction error}
	\end{subfigure}
	\hfill
	\begin{subfigure}{0.48\textwidth}
		\centering
		\includegraphics[width=\textwidth]{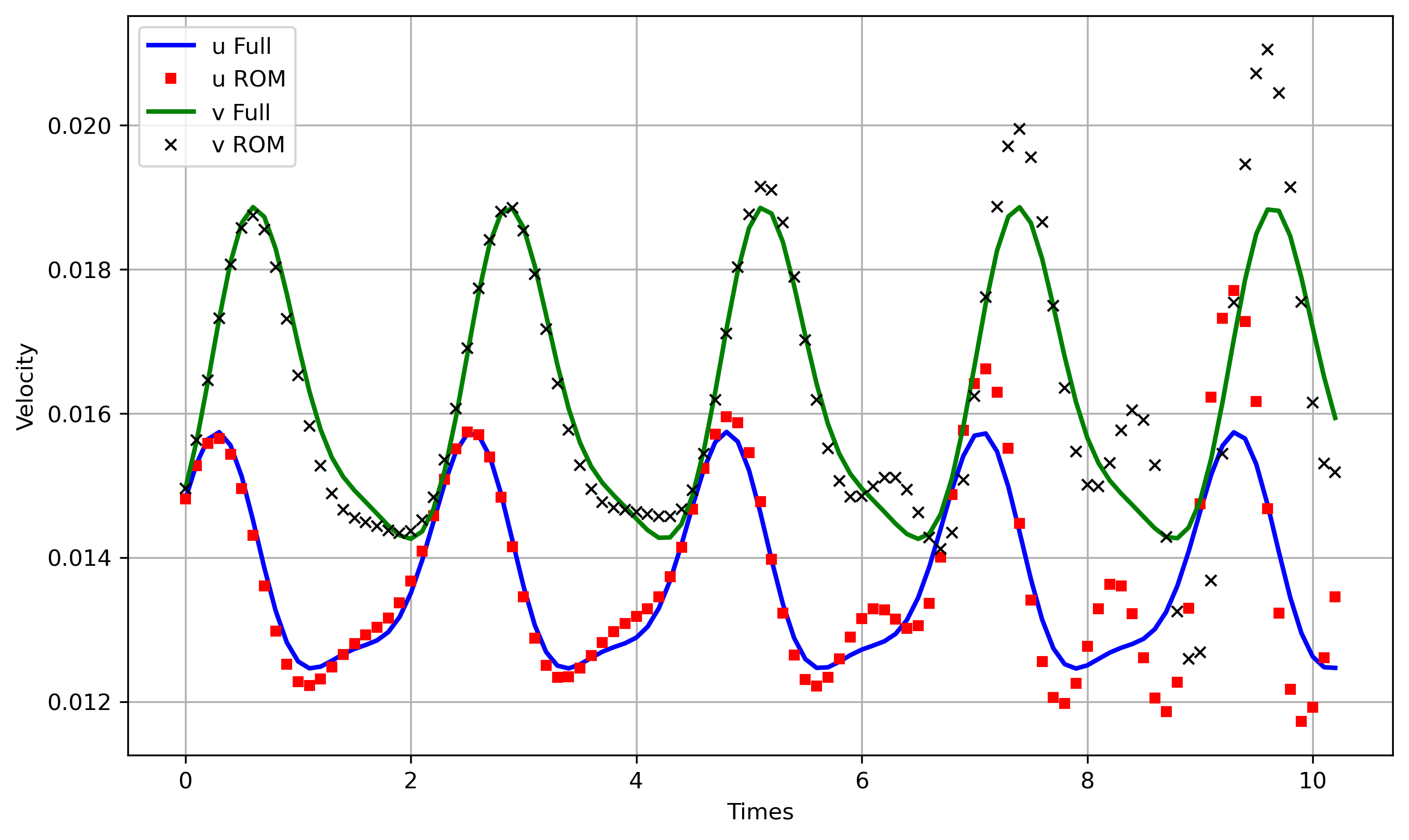}
		\caption{\small Probe 1 -- $Re=9750$ }
	\end{subfigure}
	
	\vspace{0.5cm}
	
	\begin{subfigure}{0.48\textwidth}
		\centering
		\includegraphics[width=\textwidth]{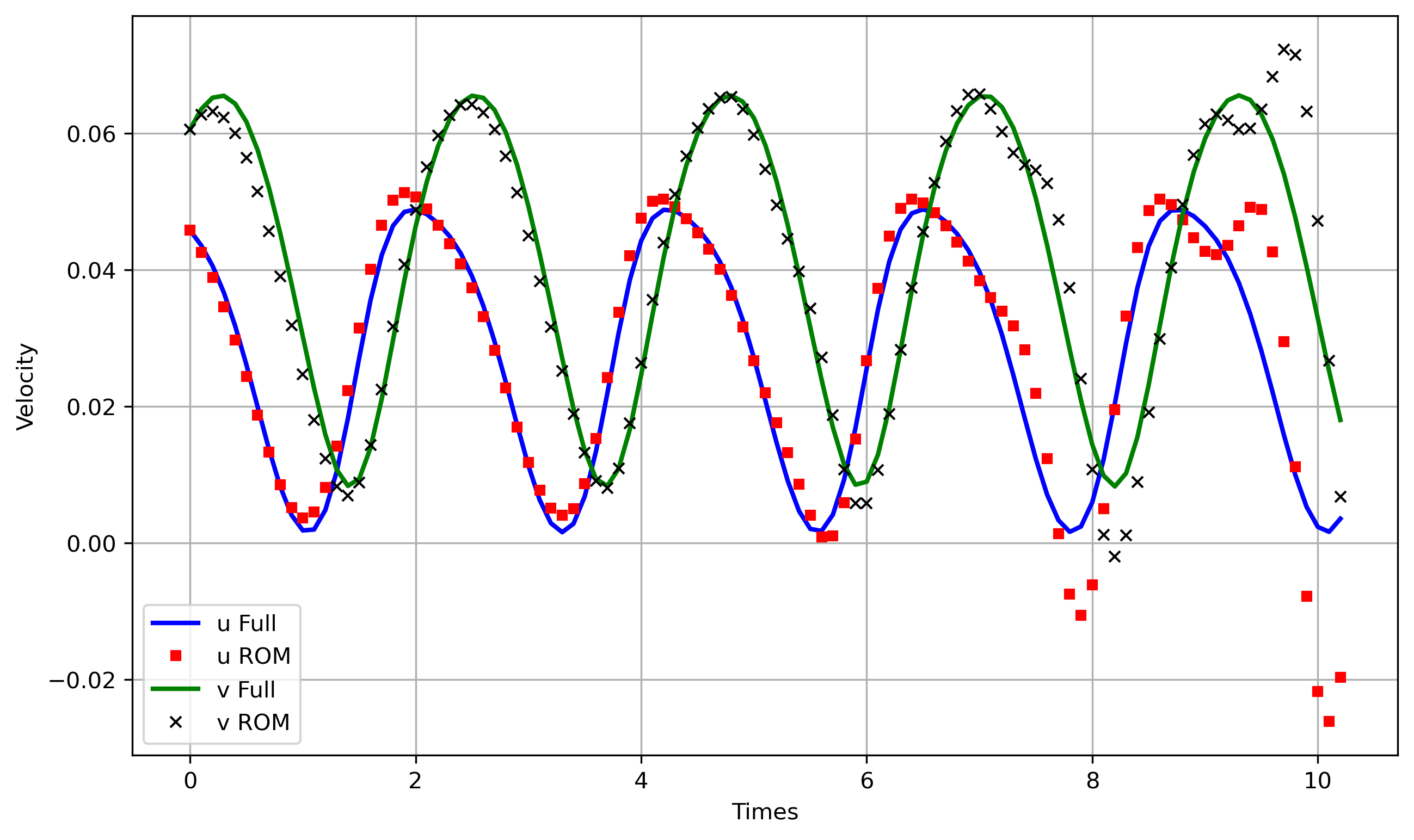}
		\caption{\small Probe 2 -- $Re=9750$ }
	\end{subfigure}
	\hfill
	\begin{subfigure}{0.48\textwidth}
		\centering
		\includegraphics[width=\textwidth]{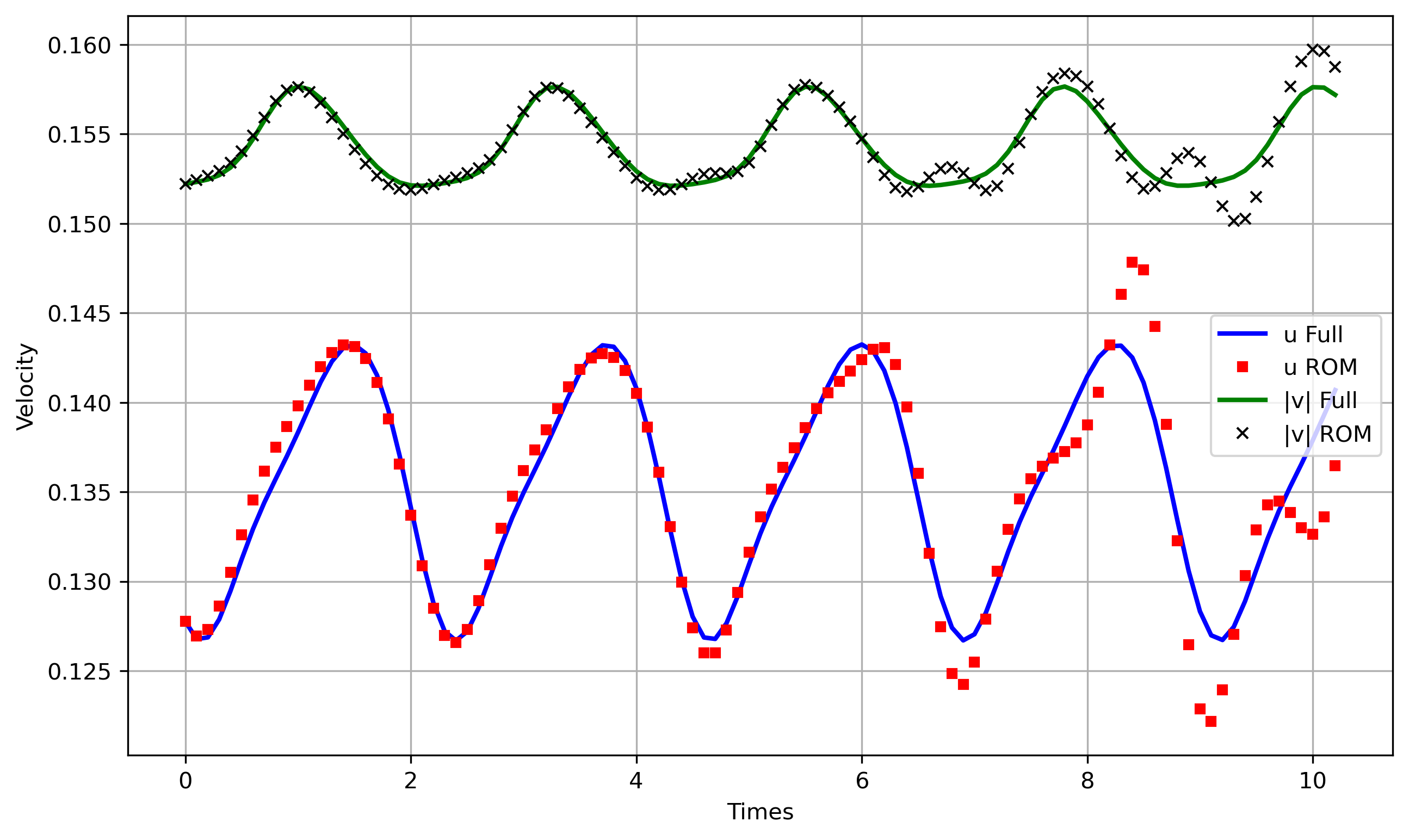}
		\caption{\small Probe 3 -- $Re=9750$ }
	\end{subfigure}
	
	\caption{Reconstruction error and velocity components (horizontal $u$, vertical $v$) measured at probes for Case E (sampling $Re \in \{8750, 9000, 9250, 9500\}$).}
	\label{fig:lid_cavity_caseE}
\end{figure}

To provide a more detailed assessment of the reconstruction results, the temporal evolution of the relative reconstruction error and the temporal evolution of the velocity components at the three probe locations are considered. As mentioned previously, the probes are located in the corners of the cavity, where the flow dynamics are more complex, providing a more demanding assessment of the reconstruction accuracy. To maintain the readability of the manuscript, only the results corresponding to the configuration with the lowest errors, Case C (Figure \ref{fig:lid_cavity_caseC}), and the configuration with the highest errors, Case E (Figure \ref{fig:lid_cavity_caseE}), are presented.

Figure \ref{fig:lid_cavity_caseC} a) shows the temporal evolution of the relative reconstruction error for the Reynolds numbers included in the training set and for the interpolated test case, $Re=9250$. As expected, the reconstruction error increases slightly with time. For the Reynolds numbers used to construct the ROM, the maximum relative error remains below $0.04\%$, indicating that the ROM accurately represents the training data. For the interpolated case, $Re=9250$, the error is higher, as expected, but remains below $0.43\%$ over the considered time interval. Regarding the velocity reconstruction at the probe locations (figures \ref{fig:lid_cavity_caseC} b,c,d)), good agreement is observed between the reconstructed and reference temporal evolutions of both the horizontal and vertical velocity components at probes 1, 2, and 3. The differences remain small over most of the time interval, with some discrepancies appearing towards the end of the simulation. These results show that the proposed approach provides accurate velocity reconstructions even at local locations characterized by more complex flow dynamics.

For Case E, the test Reynolds number, $Re=9750$, lies above the range of Reynolds numbers used to construct the ROM and therefore corresponds to an extrapolation case. As shown in Figure \ref{fig:lid_cavity_caseE}, the reconstruction error is higher than that obtained for Case C, which corresponds to an interpolation case. For the Reynolds numbers included in the training set, the ROM maintains a low reconstruction error, with a maximum error of approximately $0.05\%$ at the final time. In contrast, for the extrapolated Reynolds number, the reconstruction error reaches approximately $3\%$ at the end of the considered time interval. A similar behavior is observed for the velocity reconstruction at probe locations 1, 2, and 3. The reconstructed velocity components remain in good agreement with the reference solutions during the first half of the time interval, with only small discrepancies observed up to approximately $t=6$ s. Beyond this time, the differences progressively increase, leading to more noticeable deviations during the second half of the simulation. This indicates that the ROM provides an accurate reconstruction during the initial part of the extrapolation case, while its accuracy progressively deteriorates as the simulation progresses. These results highlight a limitation of the proposed approach when extrapolating beyond the range of Reynolds numbers used to construct the ROM.

~\\[0.2cm]
\paragraph{b) ROM for the pressure}~\\
For each case A–E, the ROM associated with the Poisson equation is constructed using 40 pressure POD modes and 20 velocity POD modes. As for the construction of the ROM associated with the momentum equation, only 70\% of the available extended time snapshots are retained, randomly selected. Although the velocity ROM is built using 40 velocity modes, only the predictions associated with the first 20 modes are considered. Indeed, since the pressure ROM is based on coefficients derived from the velocity ROM, small errors in the velocity reconstruction can lead to significant errors in the pressure field. It is therefore preferable to retain fewer modes with accurate predictive capability rather than include additional higher-order modes that may act as noise.

\begin{table}[htbp]
	\centering
	\begin{tabular}{cccccc}
		\toprule
		Training& Test Re & Mean abs. & Max abs. & Mean $\%$ & Max $\%$ \\
		\midrule
		\rowcolor{lightgray}
		Case A &8750 & $4.32 \times 10^{-4}$ & $8.13 \times 10^{-4}$ & $6.53$ & $12.39$ \\
		Case B &9000 & $1.51 \times 10^{-5}$ & $3.97 \times 10^{-5}$ & $2.29 \times 10^{-1}$ & $5.99 \times 10^{-1}$ \\
		\rowcolor{lightgray}
		Case C &9250 & $1.42 \times 10^{-5}$ & $3.78 \times 10^{-5}$ & $2.18 \times 10^{-1}$ & $5.81 \times 10^{-1}$ \\
		Case D &9500 & $3.65 \times 10^{-5}$ & $7.89 \times 10^{-5}$ & $5.56 \times 10^{-1}$ & $1.19$ \\
		\rowcolor{lightgray}
		Case E &9750 & $1.38 \times 10^{-3}$ & $2.81 \times 10^{-3}$ & $2.11 \times 10^{1}$ & $43.80$ \\
		\bottomrule
	\end{tabular}
	\caption{Pressure reconstruction errors (mean absolute error, max absolute error, mean percentage error, and max percentage error) for different sampling cases used to construct the ROM. Cases A and E correspond to extrapolation settings, whereas Cases B--D correspond to interpolation settings.}
	\label{tab:rec_error_pressure}
\end{table}
Table \ref{tab:rec_error_pressure} presents the mean error and maximum error for the pressure reconstruction in Cases A--E. As observed for the velocity reconstruction, the interpolation cases B--D yield substantially lower errors than the extrapolation cases A and E. For the interpolation cases, the mean relative error remains below $0.56\%$, with values of $0.229\%$, $0.218\%$, and $0.556\%$ for Cases B, C, and D, respectively. The corresponding maximum relative errors remain below $1.2\%$. These results indicate that the ROM provides a good pressure reconstruction for Reynolds numbers within the range used to construct the model. In contrast, a significant increase in the reconstruction error is observed for the extrapolation cases. For Case A, the mean relative error reaches $6.53\%$, with a maximum relative error of $12.39\%$, while for Case E, these errors increase to $21.1\%$ and $43.8\%$, respectively. These results show that the proposed ROM provides accurate pressure reconstructions for interpolation cases, whereas its accuracy is substantially reduced when extrapolating beyond the training range.

Since the ROM does not accurately reconstruct the pressure field for the extrapolation cases (Cases A and E), the corresponding pressure time histories at the probe locations are not presented. Only the temporal evolution of the pressure reconstruction error and the pressure time histories at the probe locations are shown for Cases C and D, corresponding to the interpolation cases with the lowest and highest reconstruction errors, respectively.
\begin{figure}[!htbp]
	\centering
	
	\begin{subfigure}{0.48\textwidth}
		\centering
		\includegraphics[width=\textwidth]{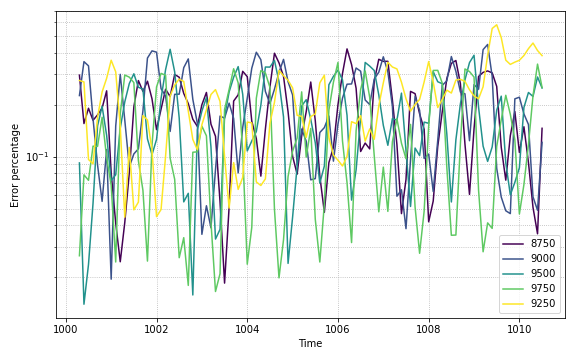}
		\caption{\small Reconstruction error}
	\end{subfigure}
	\hfill
	\begin{subfigure}{0.48\textwidth}
		\centering
		\includegraphics[width=\textwidth]{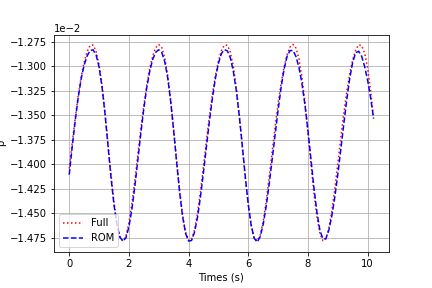}
		\caption{\small Probe 1 -- $Re=9250$ }
	\end{subfigure}
	
	\vspace{0.5cm}
	
	\begin{subfigure}{0.48\textwidth}
		\centering
		\includegraphics[width=\textwidth]{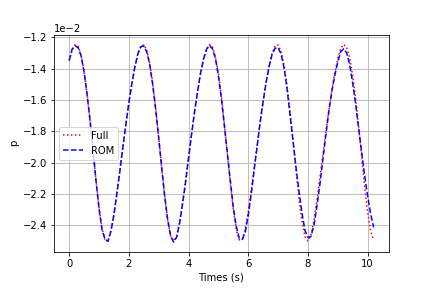}
		\caption{\small Probe 2 -- $Re=9250$ }
	\end{subfigure}
	\hfill
	\begin{subfigure}{0.48\textwidth}
		\centering
		\includegraphics[width=\textwidth]{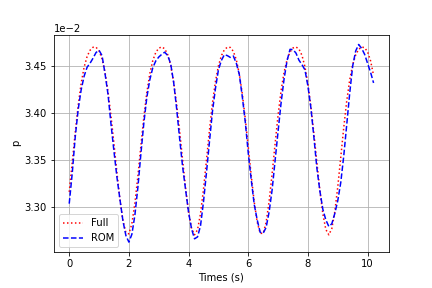}
		\caption{\small Probe 3 -- $Re=9250$ }
	\end{subfigure}
	
	\caption{Reconstruction error and pressured measured at probes for Case C (sampling $Re \in \{8750, 9000, 9500, 9750\}$).}
	\label{fig:lid_cavity_caseC_pressure}
\end{figure}

For Case C (see Figures \ref{fig:lid_cavity_caseC_pressure}), the temporal evolution of the pressure reconstruction error for the test Reynolds number, $Re=9250$, remains below $0.5\%$ and is comparable to that obtained for the Reynolds numbers included in the sampling set. The error also remains relatively constant over time. At probe locations 1, 2, and 3, the reconstructed pressure time histories are in good agreement with the reference solutions, with only small discrepancies observed, mainly around the pressure peaks.

\begin{figure}[!htbp]
	\centering
	\begin{subfigure}{0.48\textwidth}
		\centering
		\includegraphics[width=\textwidth]{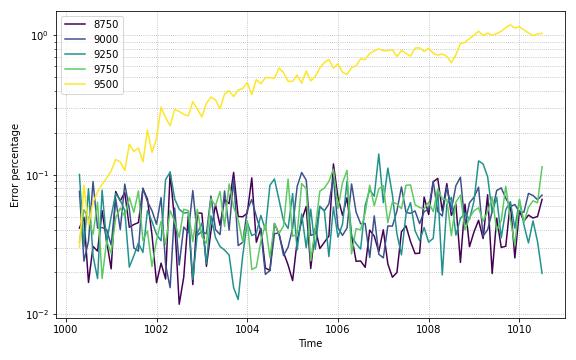}
		\caption{\small Reconstruction error}
	\end{subfigure}
	\hfill
	\begin{subfigure}{0.48\textwidth}
		\centering
		\includegraphics[width=\textwidth]{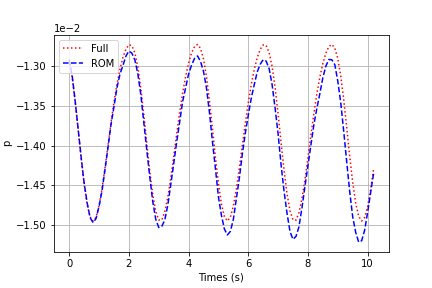}
		\caption{\small Probe 1 -- $Re=9500$ }
	\end{subfigure}
	
	\vspace{0.5cm}
	
	\begin{subfigure}{0.48\textwidth}
		\centering
		\includegraphics[width=\textwidth]{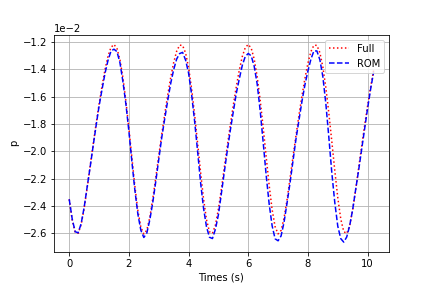}
		\caption{\small Probe 2 -- $Re=9500$ }
	\end{subfigure}
	\hfill
	\begin{subfigure}{0.48\textwidth}
		\centering
		\includegraphics[width=\textwidth]{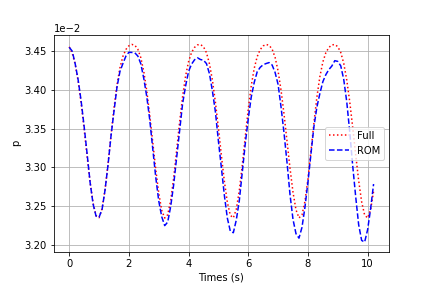}
		\caption{\small Probe 3 -- $Re=9500$ }
	\end{subfigure}
	
	\caption{Reconstruction error and pressure measured at probes for Case D (sampling $Re \in \{8750, 9000, 9250, 9750\}$).}
	\label{fig:lid_cavity_caseD_pressure}
\end{figure}
For Case D (see Figures \ref{fig:lid_cavity_caseD_pressure}) the reconstruction is less accurate. For the test Reynolds number, $Re=9500$, the pressure reconstruction error is approximately $1\%$, compared with around $0.1\%$ for the Reynolds numbers included in the sampling set. The pressure time histories at probe locations 1, 2, and 3 remain in overall good agreement with the reference solutions, although larger discrepancies are observed than in Case C. These discrepancies are also more pronounced than those observed for the velocity reconstruction. This difference can be attributed to the dependence of the pressure reconstruction on the reconstructed velocity field, since errors in the velocity reconstruction can propagate into the pressure prediction and accumulate over the course of the simulation.  In addition, these discrepancies are higher than for the velocities, because the velocity errors are introduced into the reduced order model for the pressure.

\paragraph{c) ROM for the pressure with the exact velocity coefficients}~\\
As previously observed, the ROM associated with the Poisson equation performs poorly in extrapolation. In order to determine whether this error originates from an inaccurate model identification or from the velocity approximation itself, we solve the pressure ROM using the coefficients obtained by projecting the velocity field from the full order model  onto the velocity POD basis. This procedure provides the exact velocity coefficients.

\begin{table}[!htbp]
	\centering
	\begin{tabular}{cccccc}
		\toprule
		Training& Test Re & Mean abs. & Max abs. & Mean $\%$ & Max $\%$ \\
		\midrule
		\rowcolor{lightgray}
		Case A &8750 & $1.10 \times 10^{-5}$ & $1.26 \times 10^{-5}$ & $1.66 \times 10^{-1}$ & $1.89 \times 10^{-1}$ \\
		Case E &9750 & $7.47 \times 10^{-5}$ & $9.68 \times 10^{-5}$ & $1.15$  & $1.48$ \\
		\bottomrule
	\end{tabular}
	\caption{Reconstruction errors (mean absolute error, max absolute error, mean percentage error and max percentage error) for sampling cases A and E by using the exact POD velocity coefficients.}
	\label{tab:rec_error_pressure_ROM_POD}
\end{table}

\begin{figure}[!htbp]
	\centering
	
	\begin{subfigure}{0.48\textwidth}
		\centering
		\includegraphics[width=\textwidth]{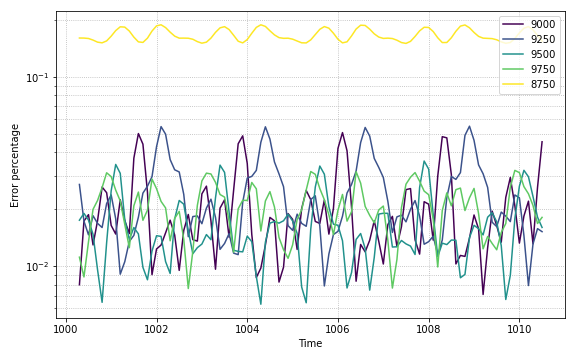}
		\caption{\small Reconstruction error}
	\end{subfigure}
	\hfill
	\begin{subfigure}{0.48\textwidth}
		\centering
		\includegraphics[width=\textwidth]{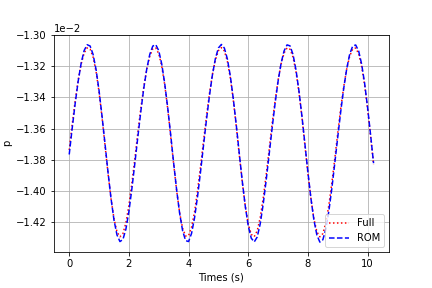}
		\caption{\small Probe 1 -- $Re=8750$ }
	\end{subfigure}
	
	\vspace{0.5cm}
	
	\begin{subfigure}{0.48\textwidth}
		\centering
		\includegraphics[width=\textwidth]{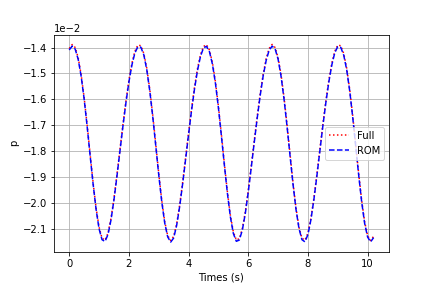}
		\caption{\small Probe 2 -- $Re=8750$ }
	\end{subfigure}
	\hfill
	\begin{subfigure}{0.48\textwidth}
		\centering
		\includegraphics[width=\textwidth]{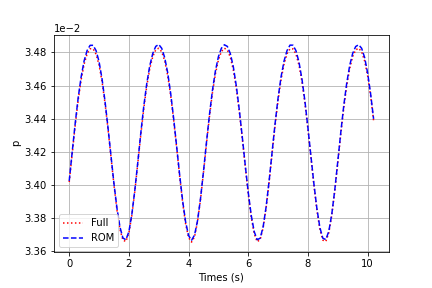}
		\caption{\small Probe 3 -- $Re=8750$ }
	\end{subfigure}
	\caption{Reconstruction error and pressure measured at probes for Case A (sampling $Re \in \{9000, 9250, 9500, 9750\}$) by using the exact POD velocity coefficients.}
	\label{fig:lid_cavity_caseA_pressure_true}
\end{figure}

This procedure enables to obtain mean absolute errors and maximum errors for the pressure reconstruction which are much smaller than when the velocity coefficients calculated
with the ROM were used (see Table \ref{tab:rec_error_pressure_ROM_POD}).For Case A ($Re=8750$), the mean relative error is $0.166\%$, while the maximum relative error is $0.189\%$. For Case E  these errors increase to $1.15\%$ and $1.48\%$, respectively. Furthermore, the time evolutions of the pressure at probe locations 1, 2 and 3 with this method are very close to the reference solution (see Figure \ref{fig:lid_cavity_caseA_pressure_true}). Predicting the pressure with the ROM and the velocity coefficients calculated with the POD yields thus accurate results. These results indicate that predicting the pressure using the ROM with the exact POD velocity coefficients provides an accurate pressure reconstruction. They confirm that the proposed approach for constructing the pressure ROM is appropriate, while its reconstruction accuracy is influenced by the accuracy of the velocity field provided by the velocity ROM.

\subsection{Fluid flow past a porous square obstacle}
\subsubsection{Studied configuration}
\begin{figure}[!h]
	\centering		
	\includegraphics[scale=0.4]{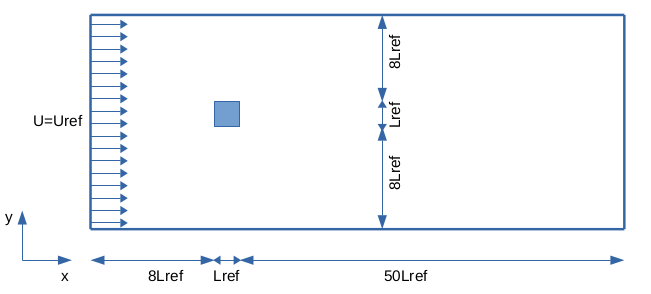} 
	\caption{Flow past a porous square obstacle}	
	\label{obstacle}	
\end{figure}

The flow past a porous square obstacle in an unconfined channel is studied here (see Figure \ref{obstacle}). The flow is unsteady, laminar and incompressible.
The computations, carried out with the solver pisoFoam of openFOAM, have been validated in a previous paper \cite{allery_beghein_2025}.
Several simulations have been performed, for various Reynolds numbers, based on the inlet velocity $U$ and the obstacle height $L_{ref}$: $Re\in\{150,160,170,180,190\}$.

Very few studies on ROMs have been devoted to flows in porous media. In these studies, the nonlinear Forchheimer term has generally been neglected, as it cannot be straightforwardly expressed in reduced form during the Galerkin projection.
In a previous work \cite{allery_beghein_2025}, the Forchheimer term was incorporated into the Galerkin projection using neural networks.
The approach considered in the present work is well suited to this type of problem and makes it possible to account for the nonlinear Forchheimer term directly within the ROM, without requiring a specific computation at each time step, as would be the case with a Galerkin projection approach.

\subsubsection{Results obtained for a fixed Reynolds number}
In this paragraph, we first consider a simulation performed for a fixed Reynolds number Re=170.
The velocity and pressure POD bases are constructed from 125 snapshots uniformly distributed over a time interval corresponding to a non-dimensional period of $T^{*}=\fract{T U}{L_{ref}}=25.4$. The snapshots are separated by a non-dimensional sampling interval of $\Delta t^{*}=0.204816$.  To obtain a more accurate approximation of the temporal derivatives of the POD coefficients, $\alpha_{\mathrm{POD}}^i$, the set of POD coefficients is enriched by means of temporal interpolation. In the present work, the derivatives are evaluated using a first-order finite-difference scheme with a non-dimensional time step of $\Delta t^{*}=0.0025$. Finally, the ROM is constructed using only $50\%$ of the coefficients from the enriched dataset, randomly selected over the non-dimensional time interval $[0,\,12.8]$.~\\

\begin{figure}[!htbp]
	\centering
	
	\begin{subfigure}{0.43\textwidth}
		\centering
		\includegraphics[width=\textwidth]{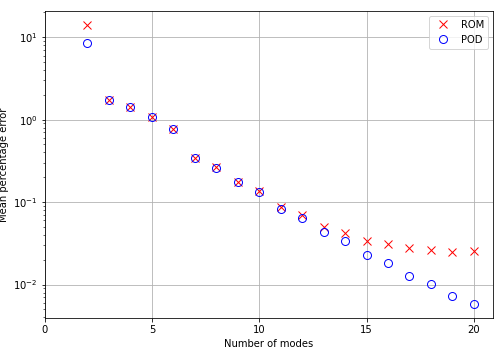}
		\caption{\small Reconstruction error}
	\end{subfigure}
	\hfill
	\begin{subfigure}{0.5\textwidth}
		\centering
		\includegraphics[width=\textwidth]{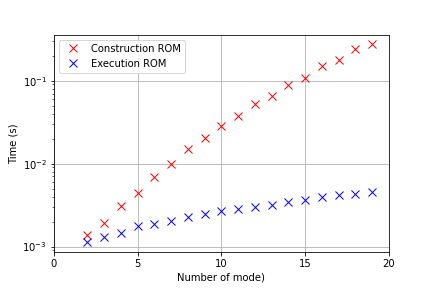}
		\caption{\small Computing time}
	\end{subfigure}
	\caption{Left: mean percentage velocity reconstruction error as a function of the number of modes for the exact POD reconstruction and the ROM. Right: computational times for ROM construction and simulation ($Re=170$).}
	\label{fig:lid_forcheimer_unparam_vitesse}
\end{figure}

\paragraph{a) ROM for the velocity}~\\
\indent
Figure \ref{fig:lid_forcheimer_unparam_vitesse} (left) compares the mean relative velocity reconstruction error obtained with the exact POD representation and with the ROM for different numbers of retained modes. The error is evaluated over the entire POD sampling interval, $[0,25.4]$, while the ROM is constructed using only the non-dimensional time interval $[0,12.8]$. As observed for the lid-driven cavity, the reconstruction error obtained with the ROM remains close to that of the exact POD representation up to 14 modes, for which the mean relative error is approximately $0.3\%$.  These results confirm that the identification approach provides an accurate velocity ROM, including when it is evaluated beyond the time interval used for its construction.

\begin{figure}[!htbp]
	\centering
	
	\begin{subfigure}{0.48\textwidth}
		\centering
		\includegraphics[width=\textwidth]{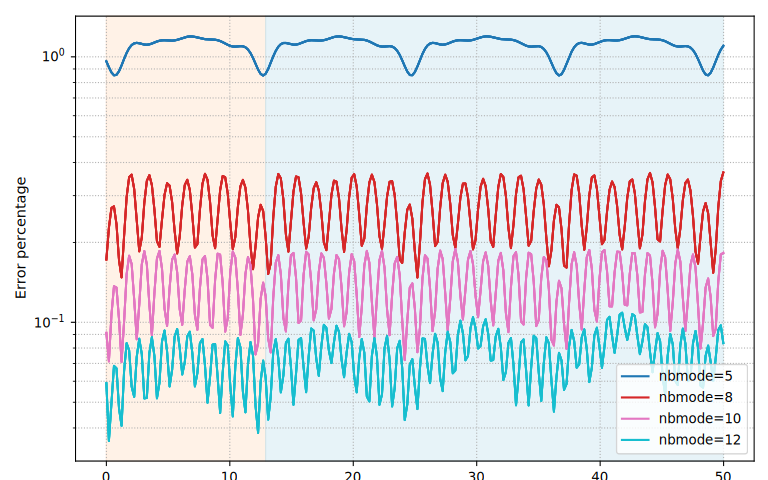}
		\caption{\small Reconstruction error --  $Re=170$}
	\end{subfigure}

	\vspace{0.5cm}
	\begin{subfigure}{0.48\textwidth}
		\centering
		\includegraphics[width=\textwidth]{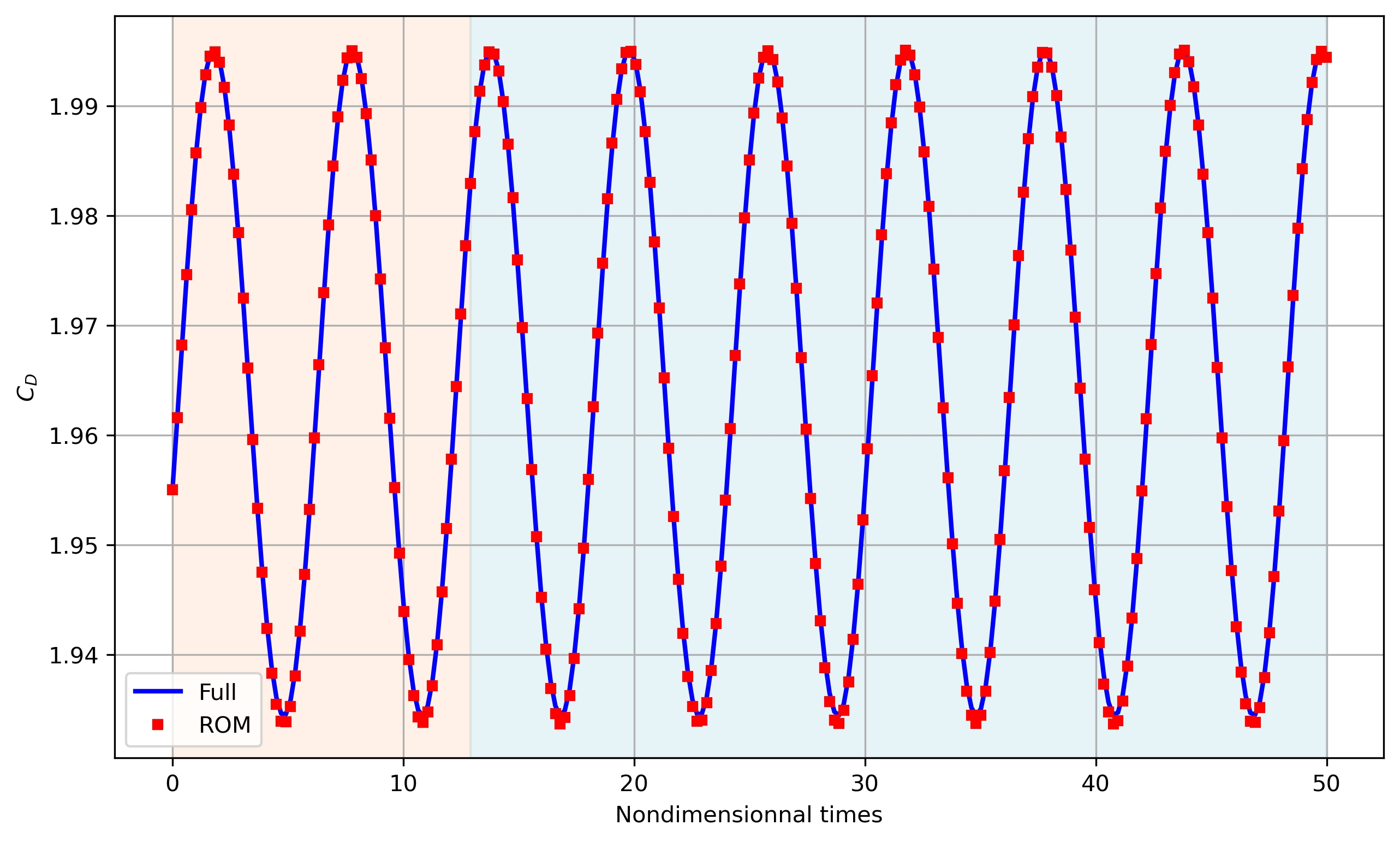}
		\caption{\small Drag coefficient -- $Re=170$ }
	\end{subfigure}
	\hfill
	\begin{subfigure}{0.48\textwidth}
		\centering
		\includegraphics[width=\textwidth]{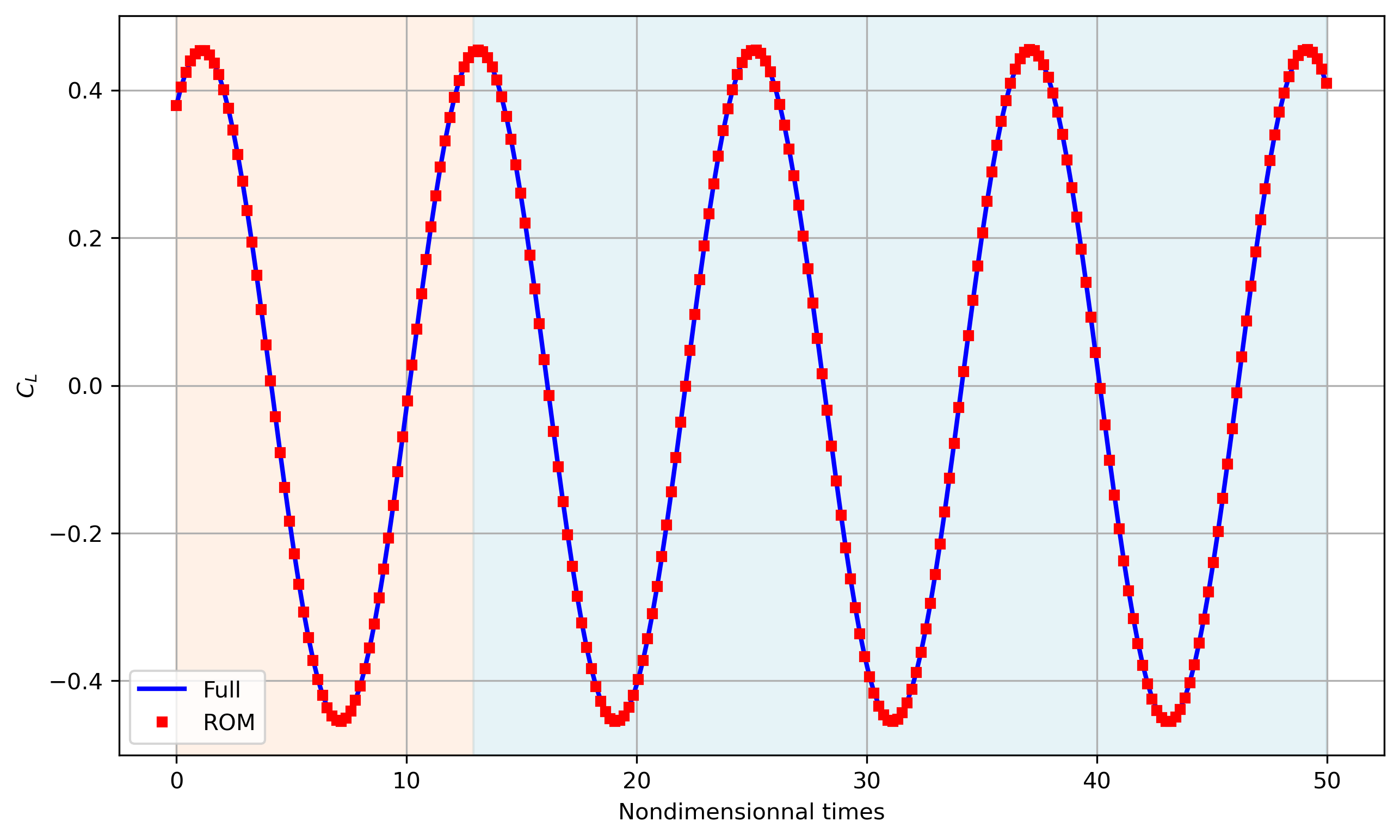}
		\caption{\small Lift coefficient -- $Re=170$}
	\end{subfigure}

	\caption{Velocity reconstruction error as a function of the number of modes, drag coefficient, and lift coefficient for $Re=170$ and $\NbmodeU=10$. Light orange background: ROM training period; light blue background: ROM prediction period.}
	\label{fig:forcheimer_velocity_unParam}
\end{figure}

\begin{figure}[!htbp]
	\centering
	\includegraphics[scale=0.4]{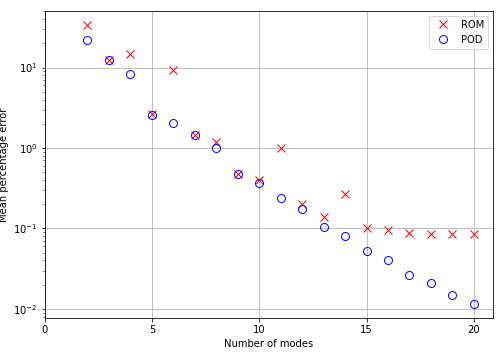}
	\caption{Mean percentage pressure reconstruction error as a function of the number of modes for the exact POD reconstruction and the ROM.}
	\label{fig:forcheimer_unparam_pression}
\end{figure}

Figure \ref{fig:forcheimer_velocity_unParam} shows the time evolutions of the reconstruction error for the velocity, and of the drag and lift coefficients. The reconstruction error decreases when the number of modes increases. In a previous work \cite{allery_beghein_2025}, we showed that the temporal coefficients exhibit a time shift relative to the reference solution when the ROM obtained by Galerkin projection is not corrected using neural networks. Here, we can see that with the temporal coefficients obtained with the identification method, the time shift disappears, there is a perfect agreement of the drag and lift coefficients obtained with the identification method, compared with the reference solution. In addition, the ROM prediction remains stable over time, even for times significantly beyond the training period.
\paragraph{b) ROM for pressure}~\\
\indent For the pressure ROM, Figure \ref{fig:forcheimer_unparam_pression} shows the evolution of the mean relative reconstruction error with the number of retained pressure POD modes. In this case, the number of velocity modes used to evaluate the right-hand side of the pressure ROM is chosen equal to the number of pressure modes retained in the model. The error is evaluated over the complete POD sampling interval, $[0,12.8]$, whereas the pressure ROM is identified from data restricted to the initial interval, $[0,25.4]$. The ROM provides a reconstruction error close to that of the exact POD representation for up to 13 modes, with a mean relative error of approximately $0.2\%$. In contrast to the velocity case, the error does not decrease monotonically as the number of modes increases, and slight increases can be observed for some mode numbers. For more than 13 modes, the error decreases only slightly.
% This results indicate that the identification procedure provides a reliable pressure ROM.

\begin{figure}[!htbp]
	\centering
	
	\begin{subfigure}{0.48\textwidth}
		\centering
		\includegraphics[width=\textwidth]{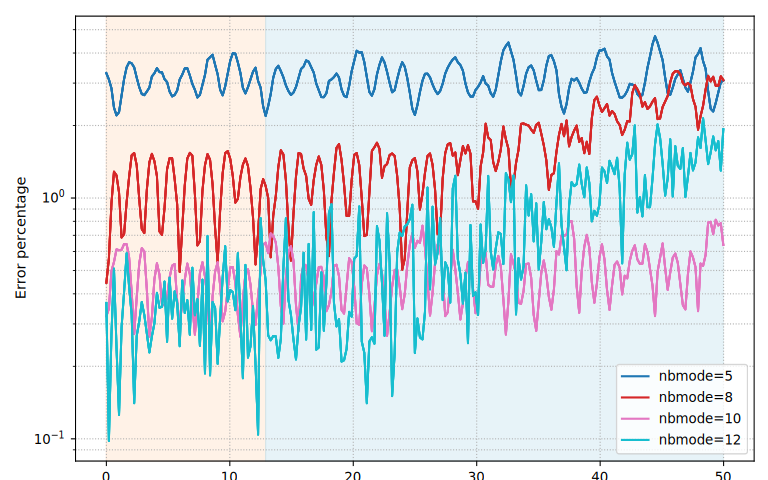}
		\caption{\small Reconstruction error}
	\end{subfigure}
	\hfill
	\begin{subfigure}{0.48\textwidth}
		\centering
		\includegraphics[width=\textwidth]{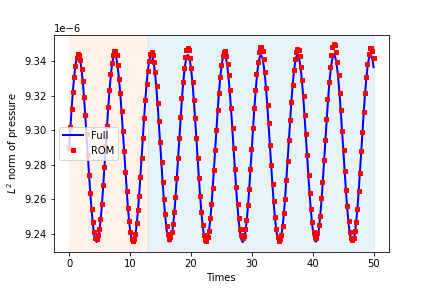}
		\caption{\small $L^2$ norm of pressure}
	\end{subfigure}

	\caption{Reconstruction error as a function of the number of modes and $L^2$ norm of pressure for $Re=170$ and $N_p=10$. Light orange background: ROM training period; light blue background: ROM prediction period.}
	\label{fig:unparam_porous_pressure}
\end{figure}

As for the velocity prediction, the pressure reconstruction error decreases as the number of retained modes increases. In addition, the $L^2$ norm of the reconstructed pressure is in good agreement with that of the exact solution (see Figure \ref{fig:unparam_porous_pressure}). These results confirm that the proposed approach can provide an accurate ROM for flows in porous media.

\subsubsection{Results obtained by the parametric ROM}
These paragraphs investigate the numerical prediction of velocity and pres-
sure with the ROMs for parametric simulations. Three cases are considered:
\begin{itemize}
	\item[-] Case A: training set $Re \in \{150, 170, 180, 190\}$, with $Re_{\text{test}} = 160$.
	\item[-] Case B: training set $Re \in \{150, 160, 180, 190\}$, with $Re_{\text{test}} = 170$.
	\item[-] Case C: training set $Re \in \{150, 160, 170, 190\}$, with $Re_{\text{test}} = 180$.
\end{itemize}
The velocity and pressure POD bases are generated, using 215 snapshots for each Reynolds number uniformly distributed over a non-dimensional time interval of $T^{*}=47$.

\paragraph{a) ROM for the velocity}~\\
\indent To accurately evaluate the temporal derivatives of the velocity POD coefficients, $\alpha_{\mathrm{POD}}^i$, required for the ROM construction, the POD coefficient database is enriched through temporal interpolation using a non-dimensional time step of $\Delta t^{*}=0.016$. For each Reynolds number, the ROM is then built using a randomly selected subset corresponding to $50\%$ of the coefficients contained in the enriched dataset over the interval $[0,\,47]$ in non-dimensional time. Finally, the ROM is constructed using the first 50 velocity POD modes.

In Table \ref{tab:rec_error_forcheimer}, the mean absolute errors and the maximum errors are reported for the velocity reconstruction for each case. It can be observed that, for Case B, where the Reynolds number lies at the center of the training set, the errors are smaller than for Cases A and C, where the Reynolds numbers are located towards the lower and upper ends of the training set, respectively. The mean relative errors range from $1.03\%$ to $1.65\%$, while the maximum relative errors range from $3.44\%$ to $6.16\%$ across the three cases. These results show that the ROM is able to accurately reconstruct the velocity data for the three cases.
\begin{table}[!htbp]
	\centering
	\begin{tabular}{cccccc}
		\toprule
		Training& Test Re & Mean abs. & Max abs. & Mean $\%$ & Max $\%$ \\
		\midrule
		\rowcolor{lightgray}
		Case A & 160 & $9.02 \times 10^{-4}$ & $3.36 \times 10^{-3}$ & $1.65 $ & $6.16 $ \\
		Case B & 170 & $5.99 \times 10^{-4}$ & $1.99 \times 10^{-3}$ & $1.03 $ & $3.44$ \\
		Case C & 180 & $9.62 \times 10^{-4}$ & $3.35 \times 10^{-3}$ & $1.57 $ & $5.44$ \\
		\bottomrule
	\end{tabular}
	\caption{Velocity reconstruction errors (mean absolute error, max absolute error, mean percentage error, and max percentage error) for different sampling cases used to construct the ROM. }
	\label{tab:rec_error_forcheimer}
\end{table}

\begin{figure}[!htbp]
	\centering
	
	\begin{subfigure}{0.48\textwidth}
		\centering
		\includegraphics[width=\textwidth]{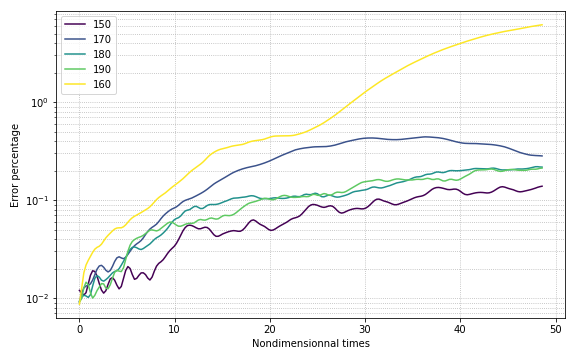}
		\caption{\small Reconstruction error --  $Re=160$}
	\end{subfigure}

	\vspace{0.5cm}
	\begin{subfigure}{0.48\textwidth}
		\centering
		\includegraphics[width=\textwidth]{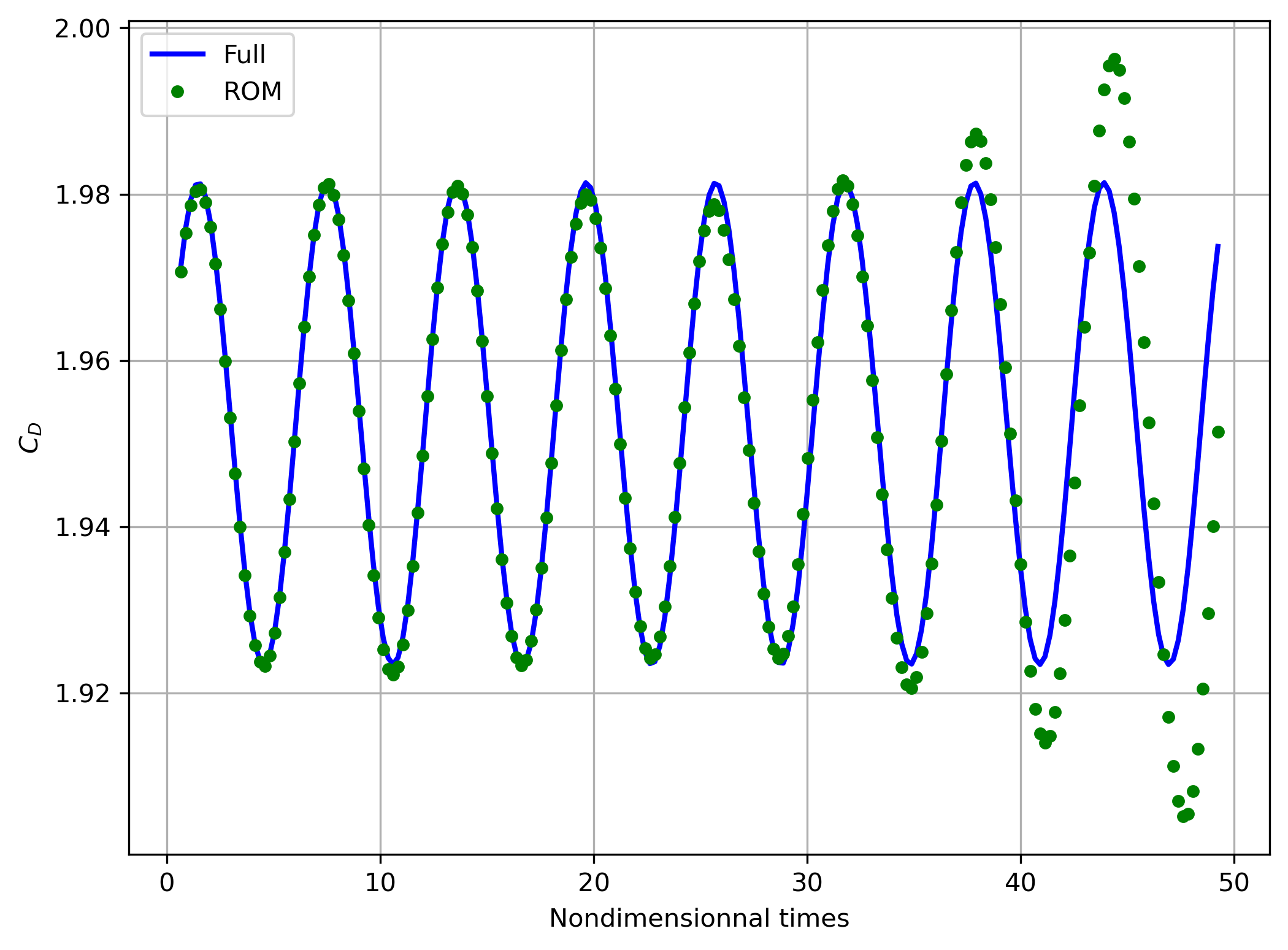}
		\caption{\small Drag coefficient -- $Re=160$ }
	\end{subfigure}
	\hfill
	\begin{subfigure}{0.48\textwidth}
		\centering
		\includegraphics[width=\textwidth]{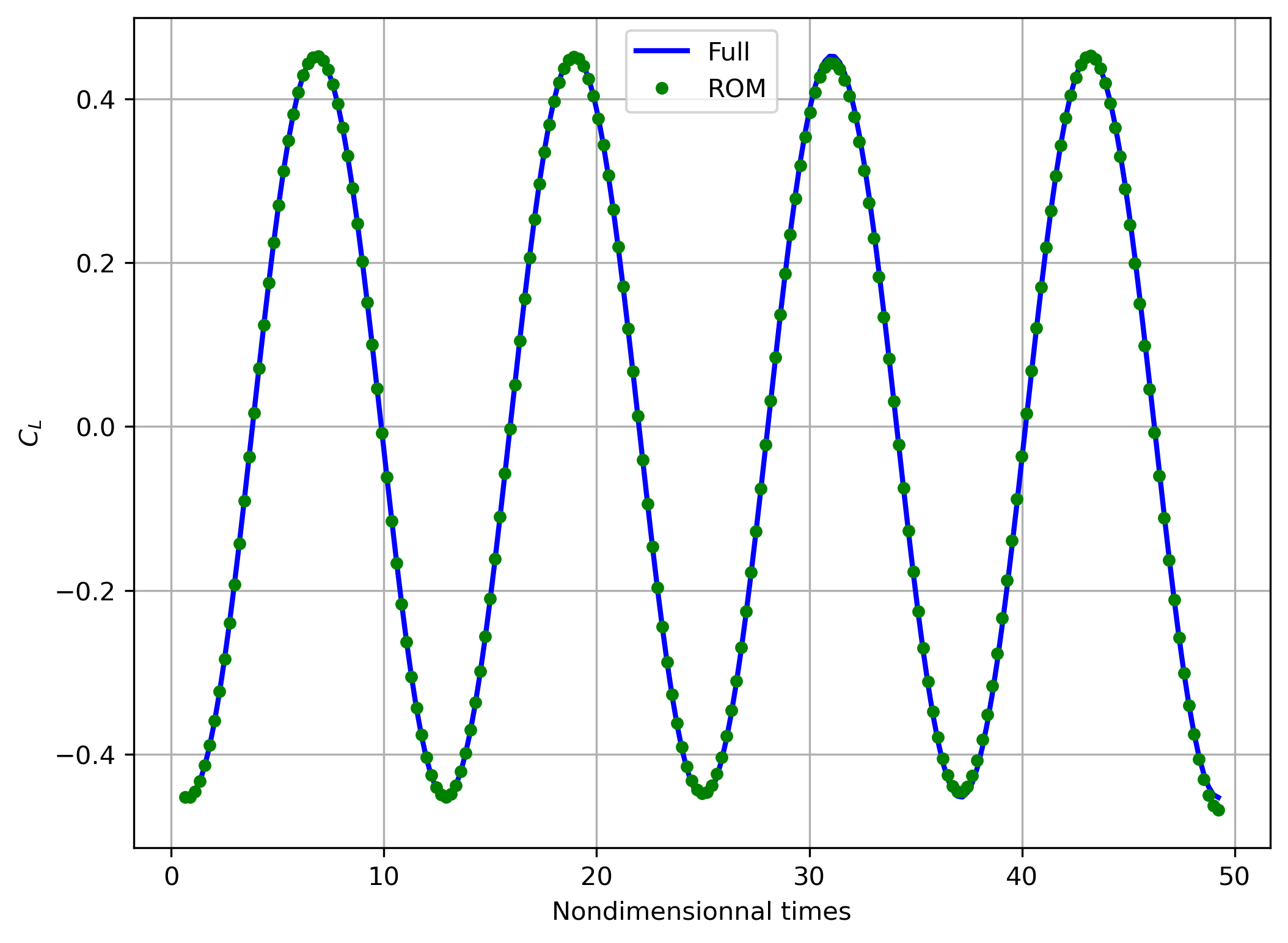}
		\caption{\small Lift coefficient -- $Re=160$ }
	\end{subfigure}

	\caption{Velocity reconstruction error, drag coefficient, and lift coefficient for Case A (sampling $Re \in \{150, 170, 180, 190\}$).}
	\label{fig:forcheimer_caseA}
\end{figure}

\begin{figure}[!htbp]
	\centering
	
	\begin{subfigure}{0.48\textwidth}
		\centering
		\includegraphics[width=\textwidth]{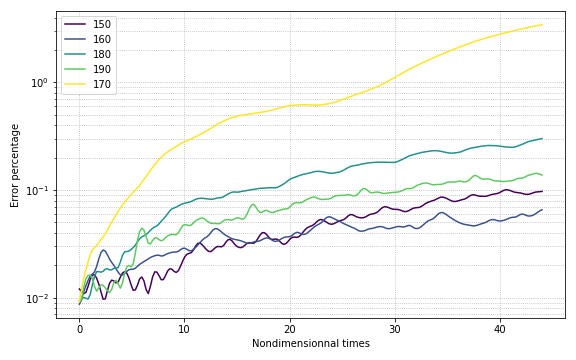}
		\caption{\small Reconstruction error --  $Re=170$}
	\end{subfigure}

	\vspace{0.5cm}
	\begin{subfigure}{0.48\textwidth}
		\centering
		\includegraphics[width=\textwidth]{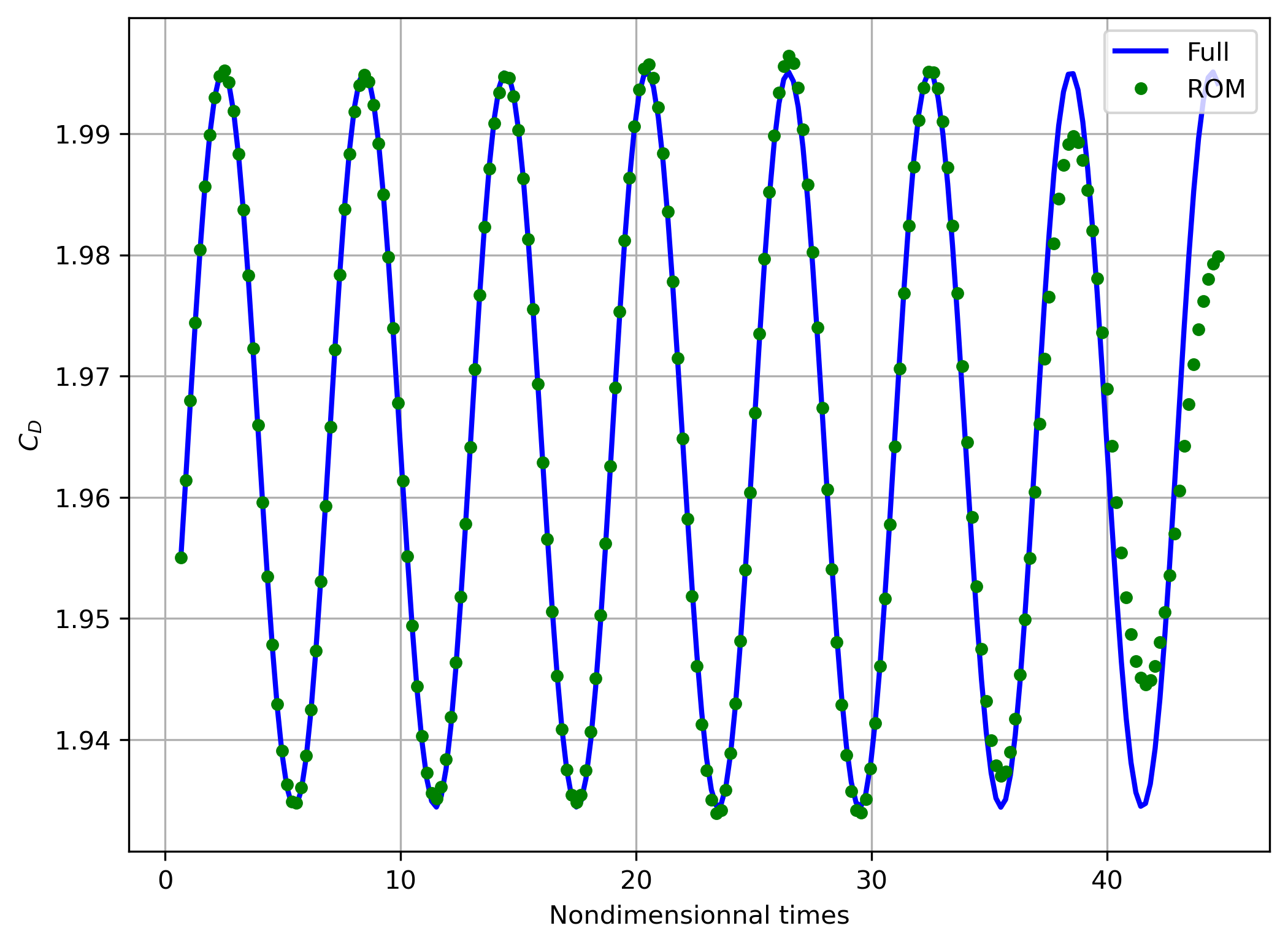}
		\caption{\small Drag coefficient -- $Re=170$ }
	\end{subfigure}
	\hfill
	\begin{subfigure}{0.48\textwidth}
		\centering
		\includegraphics[width=\textwidth]{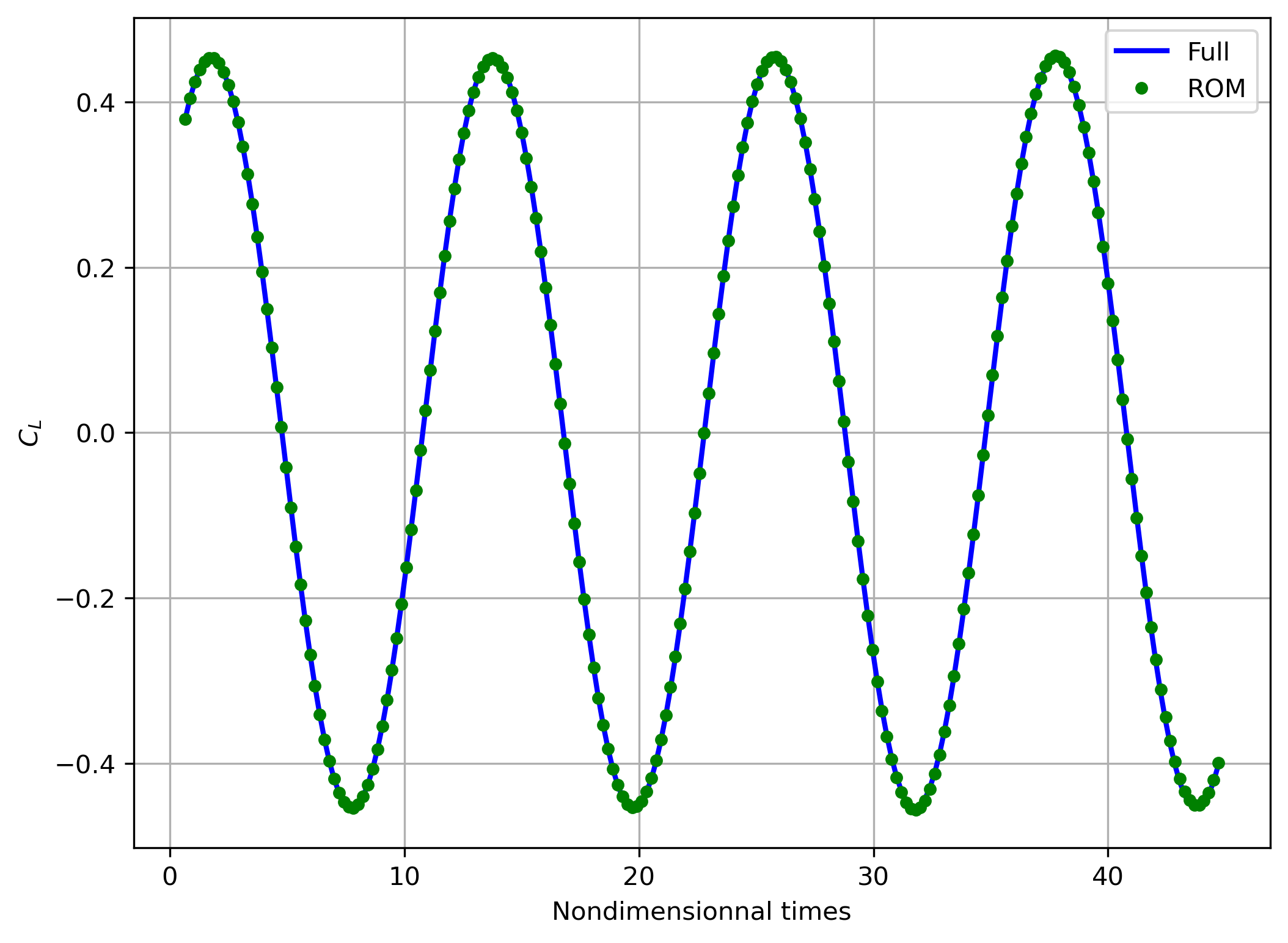}
		\caption{\small Lift coefficient -- $Re=170$ }
	\end{subfigure}

	\caption{Velocity reconstruction error, drag coefficient, and lift coefficient for Case B (sampling $Re \in \{150, 160, 180, 190\}$).}
	\label{fig:forcheimer_caseB}
\end{figure}

The temporal evolutions of the reconstruction error for Cases A and B, corresponding to the cases with the highest and lowest errors, respectively, are shown in Figures \ref{fig:forcheimer_caseA} and \ref{fig:forcheimer_caseB} a). As expected, the reconstruction error is higher for the test Reynolds number than for the Reynolds numbers included in the training set, with a maximum error of a few percent ($<6\%$), compared with a maximum error below $0.3\%$ for the training Reynolds numbers. The time evolutions of the drag and lift coefficients for cases A, B, and C are plotted in figures \ref{fig:forcheimer_caseA} b) and c), and \ref{fig:forcheimer_caseB} b) and c). The lift coefficient is well predicted by the ROMs. For the drag coefficient, differences can be observed between the reference solution and the ROM prediction. Up to approximately $t^*=35$, the drag coefficient is accurately predicted. Beyond this time, the difference progressively increases and becomes more pronounced as time increases.

\paragraph{b) ROM for the pressure}~\\
\indent The ROM associated with the Poisson equation for pressure reconstruction is constructed using $\NbmodeU = 50$ velocity modes and $\NbmodeP = 30$ pressure modes.

In Table \ref{tab:rec_error_pressure_forchheimer}, it can be noticed that for the pressure reconstruction, the same remarks can be made as for the velocity. For Cases A and C, the mean absolute error and the maximum absolute error are higher than for Case B. The mean relative error ranges from $3.34\%$ to $5.85\%$, while the maximum relative error ranges from $12.45\%$ to $26.32\%$. These relatively high errors indicate that the ROM does not provide a sufficiently accurate pressure reconstruction for this configuration, particularly at larger times.
\begin{table}[htbp]
	\centering
	\begin{tabular}{cccccc}
		\toprule
		Training& Test Re & Mean abs. & Max abs. & Mean $\%$ & Max $\%$ \\
		\midrule
		\rowcolor{lightgray}
		Case A & 160 & $4.732 \times 10^{-7}$ & $2.139 \times 10^{-6}$ & $5.85$ & $26.32$ \\
		Case B & 170 & $3.104 \times 10^{-7}$ & $1.163 \times 10^{-6}$ & $3.34$ & $12.45$ \\
		\rowcolor{lightgray}
		Case C & 180 & $5.483 \times 10^{-7}$ & $2.483 \times 10^{-6}$ & $5.18$ & $23.43$ \\
		\bottomrule
	\end{tabular}
	\caption{Pressure reconstruction errors (mean absolute error, max absolute error, mean percentage error, and max percentage error) for different sampling cases used to construct the ROM ($\NbmodeP=30$, $\NbmodeU=50$).}
	\label{tab:rec_error_pressure_forchheimer}
\end{table}

\begin{figure}[!htbp]
	\centering
	
	\begin{subfigure}{0.48\textwidth}
		\centering
		\includegraphics[width=\textwidth]{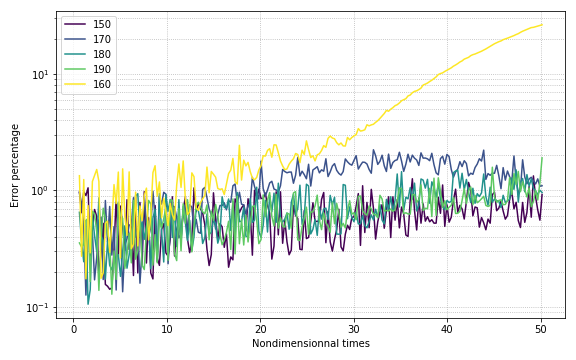}
		\caption{\small Reconstruction error $Re=160$}
	\end{subfigure}
	\hfill
	\begin{subfigure}{0.48\textwidth}
		\centering
		\includegraphics[width=\textwidth]{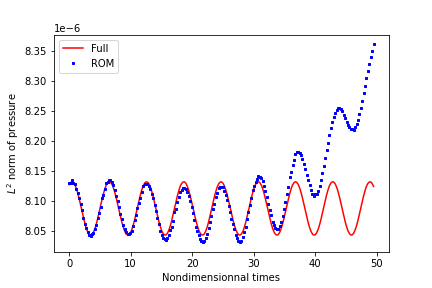}
		\caption{\small $L^2$ norm of pressure errors}
	\end{subfigure}
	\caption{Reconstruction error and $L^2$ norm of pressure errors for Case A (sampling $Re \in \{150, 170, 180, 190\}$)  and $\NbmodeP=30$, $\NbmodeU=50$.}
	\label{fig:forcheimer_caseA_pressureA}
\end{figure}

\begin{figure}[!htbp]
	\centering
	
	\begin{subfigure}{0.48\textwidth}
		\centering
		\includegraphics[width=\textwidth]{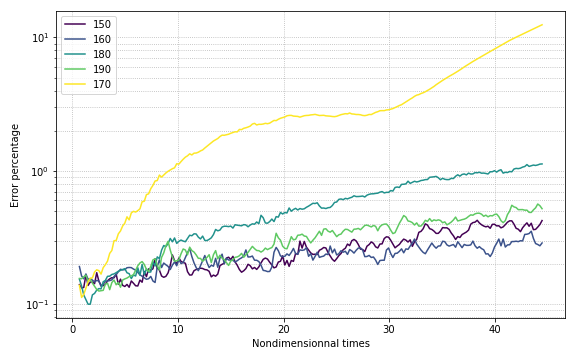}
		\caption{\small Reconstruction error $Re=170$}
	\end{subfigure}
	\hfill
	\begin{subfigure}{0.48\textwidth}
		\centering
		\includegraphics[width=\textwidth]{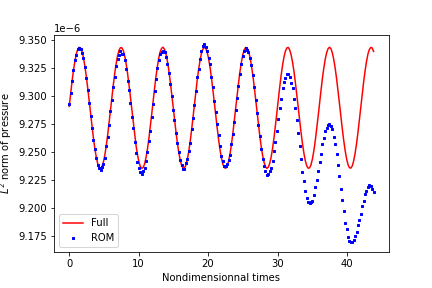}
		\caption{\small $L^2$ norm of pressure errors }
	\end{subfigure}
	\caption{Reconstruction error and $L^2$ norm of pressure errors for Case B (sampling $Re \in \{150, 160, 180, 190\}$) and $\NbmodeP=30$, $\NbmodeU=50$.}
	\label{fig:forcheimer_caseB_pressureB}
\end{figure}

The time evolutions of the reconstruction error for the pressure and of the $L^2$ norm of pressure errors for cases A and B are depicted in figures \ref{fig:forcheimer_caseA_pressureA} and \ref{fig:forcheimer_caseB_pressureB}. Since the ROM velocity coefficients are introduced into the ROM for pressure, the pressure reconstruction error is higher than that for velocity. The error remains below $2\%$ up to $t^*=30$, after which it progressively increases. A similar trend is observed for the $L^2$ norm of the pressure, with good agreement between the ROM and reference solutions up to $t^*=30$ and increasing discrepancies thereafter.

\paragraph{c) ROM for the pressure with the exact velocity coefficients}~\\

\indent The ROM associated with the Poisson equation becomes less accurate at larger times. To determine whether this error originates from the pressure ROM itself, pressure prediction is also evaluated using the exact velocity coefficients obtained by projecting the full model velocity field onto the velocity POD basis.

In table \ref{tab:rec_error_pressure_forchheimer_POD}  which presents the mean absolute error and the maximum error for the pressure, it can be remarked that the errors are smaller than when the velocity coefficients are obtained from the velocity ROM (see table \ref{tab:rec_error_pressure_forchheimer}). The mean relative error ranges from $0.11\%$ to $0.7\%$, while the maximum relative error ranges from $0.13\%$ to $1.09\%$.
\begin{table}[htbp]
	\centering
	\begin{tabular}{cccccc}
		\toprule
		Training& Test Re & Mean abs. & Max abs. & Mean $\%$ & Max $\%$ \\
		\midrule
		\rowcolor{lightgray}
		Case A & 160 & $5.6- \times 10^{-8}$ & $8.85 \times 10^{-8}$ & $6.99 \times 10^{-1}$ & $1.09$ \\
		Case B & 170 & $1.05 \times 10^{-8}$ & $1.24 \times 10^{-8}$ & $1.13 \times 10^{-1}$ & $1.32 \times 10^{-1}$ \\
		\rowcolor{lightgray}
		Case C & 180 & $1.45 \times 10^{-8}$ & $2.01 \times 10^{-8}$ & $1.41 \times 10^{-1}$ & $1.89 \times 10^{-1}$ \\
		\bottomrule
	\end{tabular}
	\caption{Pressure reconstruction errors (mean absolute error, max absolute error, mean percentage error, and max percentage error) for different sampling cases used to construct the ROM ($\NbmodeP=30$, $\NbmodeU=50$).}
	\label{tab:rec_error_pressure_forchheimer_POD}
\end{table}

\begin{figure}[!htbp]
	\centering
	
	\begin{subfigure}{0.48\textwidth}
		\centering
		\includegraphics[width=\textwidth]{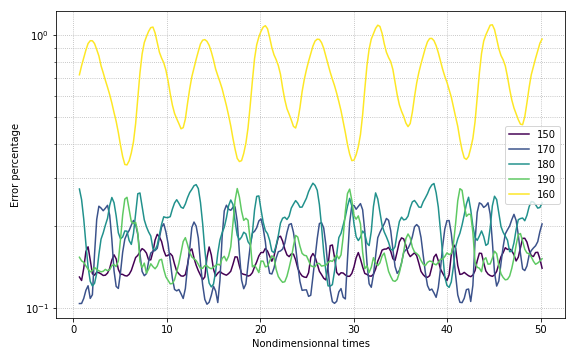}
		\caption{\small Reconstruction error $Re=160$}
	\end{subfigure}
	\hfill
	\begin{subfigure}{0.48\textwidth}
		\centering
		\includegraphics[width=\textwidth]{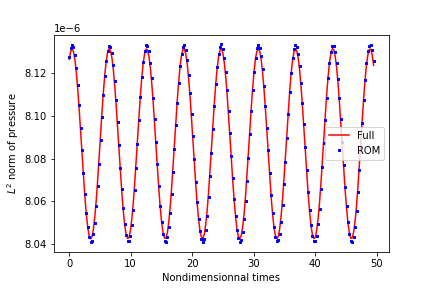}
		\caption{\small $L^2$ norm of pressure errors }
	\end{subfigure}
	\caption{Reconstruction error and $L^2$ norm of pressure errors for Case A (sampling $Re \in \{150, 170, 180, 190\}$)  and $\NbmodeP=30$, $\NbmodeU=50$.}
	\label{fig:forcheimer_caseA_pressure}
\end{figure}

\begin{figure}[!htbp]
	\centering
	
	\begin{subfigure}{0.48\textwidth}
		\centering
		\includegraphics[width=\textwidth]{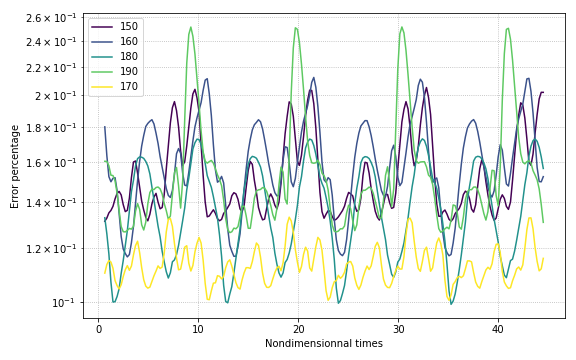}
		\caption{\small Reconstruction error $Re=170$}
	\end{subfigure}
	\hfill
	\begin{subfigure}{0.48\textwidth}
		\centering
		\includegraphics[width=\textwidth]{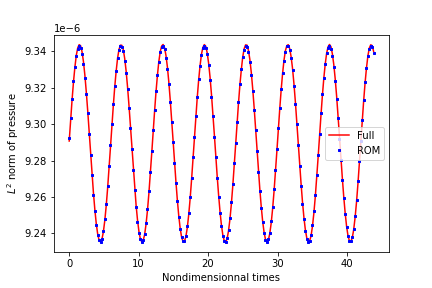}
		\caption{\small $L^2$ norm of pressure errors }
	\end{subfigure}
	\caption{Reconstruction error and $L^2$ norm of pressure errors for Case B (sampling $Re \in \{150, 160, 180, 190\}$) and $\NbmodeP=30$, $\NbmodeU=50$.}
	\label{fig:forcheimer_caseB_pressure}
\end{figure}

The reconstruction error with respect to time is also much smaller than that obtained when the ROM for pressure uses the ROM velocity coefficients (see figures \ref{fig:forcheimer_caseA_pressure} a) and \ref{fig:forcheimer_caseB_pressure} a)). It should be noted that, in this case, the error does not increase with time but instead exhibits oscillations. For Case B, the error for the test Reynolds number, $Re=170$, is smaller than that obtained for the Reynolds numbers used to construct the ROM.
In addition, the $L^2$ norm of the pressure predicted by the ROM for Cases A and B is in close agreement with the reference solution see figures \ref{fig:forcheimer_caseA_pressure} a) and  \ref{fig:forcheimer_caseB_pressure} a). These results confirm the observations made for the lid-driven cavity, namely that the proposed approach for constructing the pressure ROM is appropriate, but that its accuracy is highly dependent on the accuracy of the velocity field provided by the velocity ROM.

\section{Conclusion and perspectives}
This work presented a physics based non intrusive reduced order modeling approach for transient incompressible flows, based on the identification of reduced operators directly from temporal and parametric POD coefficients. The identification procedure relies on a sequence of uncoupled least squares problems, providing an efficient alternative to the explicit projection of the governing equations. The methodology was first assessed for the lid driven cavity flow in both monoparametric and multiparametric settings.
In the monoparametric case, the ROM accurately extrapolates the velocity and pressure fields beyond the time interval used for its construction, while remaining stable over time. The multiparametric results showed accurate predictions for both velocity and pressure at Reynolds numbers within the training range, with larger errors observed for extrapolation, particularly for the pressure field. The results also highlighted the influence of the velocity prediction on pressure reconstruction, since the pressure ROM relies on the reduced velocity coefficients.
The approach was then applied to transient flow through a porous medium with a square obstacle, including the nonlinear Forchheimer term. In the monoparametric case, the ROM provided also accurate and stable predictions of the velocity and pressure fields, as well as of the drag and lift coefficients, over time intervals extending beyond the training period. The multiparametric results also demonstrated the ability of the approach to reconstruct the velocity and pressure fields for different Reynolds numbers.
These results show that the proposed methodology can handle both the temporal and parametric variations of porous media flows while avoiding the explicit treatment of the nonlinear Forchheimer term required by a standard Galerkin projection.

\bibliographystyle{plain}
\bibliography{bibliographie_publi_Opinf}

\end{document}